\documentclass{aa}  
\usepackage{graphicx}
\usepackage{txfonts}
\begin{document}

   \title{The ambiguous transient ASASSN-17hx}

   \subtitle{A possible nova-impostor} % or a very peculiar nova

   \author{Elena Mason\inst{1}
          \and
          Steven N. Shore\inst{2}
          \and
          Paul Kuin\inst{3}
          \and
          Terry Bohlsen\inst{4}
          }

   \institute{INAF-OATS,
              Via G.B. Tiepolo 11, 34143 Trieste, Italy\\
              \email{elena.mason@inaf.it}
         \and
             Universit\'a di Pisa, Dipartimento di Fisica, Largo Enrico Fermi, Pisa, Italy %\\
             %\email{}
             %\thanks{}
             \and
        Mullard Space Science Laboratory, University College London, Holmbury St Mary, Dorking, Surrey RH5 6NT, UK%\\
        %\email{}
        \and
        ARAS group; Mirranook, Armidale, NSW, Australia
             }

   \date{Received September 15, 1996; accepted March 16, 1997}

% \abstract{}{}{}{}{} 
% 5 {} token are mandatory
 
  \abstract
   {}
   %{Some transients, classified as novae on the basis of their maximum magnitude or early decline spectra cast doubts about their true nature and whether nova impostors may exist.}
   {Some transients, although classified as novae based on their maximum and early decline optical spectra, cast doubts on their true nature and whether nova impostors might exist. }
   {We monitored a candidate nova which displayed a distinctly unusual light curve at maximum and early decline through optical spectroscopy (3000-10000 \AA, 500$<$R$<$100000) complemented with {\it Swift} UV and AAVSO optical photometry. We use the spectral line series to characterize the ejecta dynamics, structure, and mass. }
   {We found that the ejecta are in free ballistic expansion and structured as typical of classical novae. 
   However, their derived mass is at least an order of magnitude larger than the typical ejecta masses obtained for classical novae. Specifically, we found M$_{ej}\simeq$9$\times$10$^{-3}$ M$_\odot$ independent of the distance for a filling factor $\varepsilon$=1. By constraining the distance we derived $\varepsilon$ in the range 0.08-0.10,  giving a mass 7$\times$10$^{-4}\lesssim$ M$_{ej}\lesssim$9$\times$10$^{-4}$ M$_\odot$. The nebular spectrum, characterized by unusually strong coronal emission lines, confines the ionizing source energy to the range 20-250 eV, possibly peaking  in the range 75-100 or 75-150 eV.}
   {We link this source to other slow novae which showed similar behavior and suggest that they might form a distinct physical sub-group. They may result from a classical nova explosion occurring on a very low mass white dwarf or be impostors for an entirely different type of transient. 
   }


   \maketitle
%
%-------------------------------------------------------------------

\section{Introduction}
%Type Ia supernova (SN) light curves are well known standard candles, although only the fraction of "well behaved" SN-Ia define what is standard and thus the distance ladder (Mannucci private communication). For decades, classical novae (CNe) light curves have been analyzed with the same goal. While their use as standard candles is still debated (possibly because the community has not considered yet discarding the "not well behaving" novae; e.g. Selvelli and Gilmozzi 2019, Cao et al. 2012), a number of classification schemes based on  either the visual light curve rate of decline (Payne-Gaposchkin 1957) or morphology (Duerbeck 1981; Strope et al. 2010) have been proposed.  The existing collections of light curves show that they can differ widely. 
%However, a light curve alone (even in multi bands) hardly suffices to encompass the underlying physics. Most information can be grasped through high resolution spectroscopy with the widest possible spectral coverage (ideally close to panchromatic, i.e. covering the bulk of the transient emission), and extended for long times after the discovery. 

ASASSN-17hx  (also called nova Sct 2017; 17hx hereafter) was discovered on Jun 19.41 UT as a candidate bright nova (Stanek et al. 2017). Spectroscopic confirmation came one day after the announcement (Kurtenkov et al. 2017a) together with its CTIO spectral classification (Williams et al. 1991, 1994). Reports of the 17hx light curve and spectral variations filled the ATels for about 3 months (Williams \& Darnley, 2017; Saito et al. 2017; Berardi et al. 2017; Munari et al. 2017a,b,c; Pavana et al. 2017; Kuin et al. 2017; Kurtenkov et al. 2017b; Guarro et al. 2017), while, the observations continued to build a substantial data set. % in order to make sense of the transient. 
In an attempt to make sense of the peculiar light curve of ASASSN-17hx, we collected spectra and in particular high resolution spectra for 2 years after discovery. In this paper we present those spectra and our results. 
In what follows, we make no assumptions about the classical nova (CN) nature of 17hx. We will show that a central object governs the radiative budget of the ejecta, and demonstrate that the spectra arise from a freely expanding gas in ballistic motion, as in CNe. Nevertheless, this transient is dissimilar to CNe in other significant aspects.  

%--------------------------------------------------------------------
\begin{table*}
\caption{Epoch of the VLT (Very Large Telescope)/UVES (Ultraviolet and Visual Echelle Spectrograph) observations and the adopted instrument setups. In all cases the CCD readout was "fast readout", "low gain", and un-binned. In columns 4,6 and 7, whenever two values are given, the first refers to the blue arm and the second to the red-arm. Note that the resolving power (R) of the red arm depends also on the CCD, within the 2 CCD mosaic (e.g. 95000 and 100000).  }\label{uveslog}
\centering
\begin{tabular}{lcccccc}
\hline \hline
UT date & age & UT start & exptime & inst. setup &  slit & R \\ & (d) & (hr) & (s) & & (") & \\
\hline
2017/08/17 & 56 & 01:59 & 1000 & DIC1 346+564 & 0.4/0.3   & 65000/95-100000\\
2017/08/17 & 56 & 02:21 & 500 & DIC2 437+760 & 0.4/0.3 & 65000/95-100000 \\
2017/08/23 & 64 & 02:20 & 1100 & DIC1 346+564 & 0.8 & 60000\\
2017/08/23 & 64 & 02:43 & 400/300 & DIC2 437+760 & 0.8 & 60000\\
2017/09/17 & 89 & 01:24 & 1100 & DIC1 346+564  & 0.8 & 60000\\
2017/09/17 & 89 & 01:51 & 400 & DIC2 437+760 & 0.8 & 60000\\
2017/10/30 & 132 & 00:34 & 1100 & DIC1 346+564 & 0.4/0.3 & 65000/95-100000\\
2017/10/30 & 132 & 00:59 & 400 & DIC2 437+760 & 0.4/0.3 & 65000/95-100000\\
\hline
\end{tabular}
\end{table*}

\section{Observations and data reduction}

\begin{table}[]
    \caption{The log of the observations for the ARAS spectra published in this work: the top 8 spectra are the low resolution flux calibrated spectra shown in Fig.\ref{spcseq}; the bottom 17 spectra are the higher resolution data used to produce Fig.\ref{cycle}. }\label{araslog}
    \centering
    \begin{tabular}{lcccc}
         \hline\hline
UT date & age & UT start & exptime &  R \\
 & (d) &  & (s) &  \\
 \hline
2017/07/01 & 12 & 20:45 & 9149 & 598 \\
2017/07/10 & 21 & 20:18 & 8050 & 525 \\
2017/07/31 & 42 & 19:48 & 5454 & 580 \\
2017/08/09 & 51 & 19:45 & 6618 & 580 \\
2017/08/13 & 55 & 19:45 & 6181 & 580 \\
2017/08/23 & 65 & 19:06 & 5403 & 530 \\
2017/09/04 & 77 & 18:50 & 6094 & 580 \\
2017/09/08 & 81 & 18:43 & 4362 & 580 \\
\hline
2017/07/16 & 27 & 20:20 & 13250  &  11000\\
2017/07/26 & 36 & 02:04 & 10039 & 13000 \\
2017/07/30 & 40 & 01:43 & 8506 & 12000 \\
2017/08/08 & 50 & 21:34 & 7224 & 9000 \\
2017/09/01 & 74 & 19:13 & 9630 & 9000 \\
2017/09/10 & 82 & 00:19 & 7288 & 12000 \\
2017/09/15 & 87 & 00:04 & 6074 & 13000 \\
2017/09/21 & 94 & 18:54 & 3608 & 9000 \\
2017/09/28 & 101 & 18:38 & 7220 & 9000 \\
2017/10/03 & 106 & 18:51 & 6016 & 9000 \\
2017/10/05 & 108 & 18:07 & 2404 & 9000 \\
2017/10/09 & 112 & 18:00 & 7220 & 9000 \\
2017/10/16 & 119 & 18:26 & 6026 & 9000 \\
2017/10/20 & 123 & 17:52 & 7229 & 9000 \\
2017/10/24 & 127 & 17:54 & 6017 & 9000 \\
2017/10/31 & 134 & 17:37 & 6024 & 9000 \\
2017/11/11 & 145 & 17:28 & 3609 & 9000 \\
\hline
    \end{tabular}
\end{table}

\begin{table}[]
    \caption{Log of observations for the transition stage spectra. The top two lines refer to the LISA spectra, the bottom one to the TIGRE (Telescopio Internacional de Guanajuato Robotico Espectroscopico)/HEROS (Heidelberg Extended Range Optical Spectrograph) spectrum. }
    \label{1yrspclog}
    \centering
    \begin{tabular}{lcccc}
    \hline\hline
UT date  & age & UT start & exptime & R \\
  & (d) &  & (s) &  \\
  \hline
2018/03/30  & 284 & 16:42 & 1813 & 1400 \\ %JD mid=2458208.2066
2018/05/29  & 344 & 12:06 & 1998 & 1400 \\ %JD mid=2458268.0159
  \hline
2018/09/25-26  & 462,463 & 01:20 & 7200$\times$2 &  20000\\
%  &  &  &  &  \\
\hline
    \end{tabular}
\end{table}

The full characterization of a CN or an optical transient requires long term monitoring, preferably for several years. Our spectroscopic campaign extended for 2 years. The early decline phase was fortuitously monitored by UVES (Dekker et al. 2000) at the VLT within an independent program (299.D-5043 and 0100.D-0621, PI Molaro). The UVES sequence consists of four epochs spread over two months starting from two months after discovery. They sample (see light curve in Fig.\ref{lc}) two distinct optical photometric minima and the second maximum. The instrument setups and total integration times are listed in Table\ref{uveslog}. 
The UVES data were retrieved from the archive and processed using the {\it Esorex+Gasgano}  pipeline v5.9.1. Flux calibration was achieved using the instrument response curve produced by the observatory (no dedicated spectrophotometry was obtained during the observing nights). Due to the narrow slit employed (see Table\,\ref{uveslog}) for the science and the default 5" slit adopted by the observatory for the calibration program, the UVES spectra are affected by unaccounted slit losses. 

The early decline of 17hx was intensively monitored by the ARAS group\footnote{Astronomical Ring for Access to Spectroscopy: http://www.astrosurf.com/aras/} for over five months %(from day 12 to past day 150) 
with low and mid-high resolution spectroscopy. We selected from the database the spectra having R$\geq$9000, for inspection of the H$\alpha$ profile\footnote{Many of the ARAS high resolution spectra are limited to the H$\alpha$ region.}, and the low resolution (R$<$1000) flux calibrated spectra for inspection of the spectral energy distribution (SED) and comparison with the UVES spectra. Epochs, spectral resolution and exposure times for each of the selected ARAS spectra are reported in Table\,\ref{araslog}. We note that the ARAS data are available in reduced form and that the data reduction process followed by the group is the standard procedure. 

Late spectra were taken $\sim$9 and 11 months after outburst by one of us (TB) in low-resolution mode (LISA spectrograph on a 0.28 m Celestron telescope) together with broad band photometry (B, V) and have been complemented with mid-resolution spectra taken at the Heidelberg robotic telescope TIGRE with HEROS, almost 15 months after outburst (see  Table\,\ref{1yrspclog}). Note that LISA spectra were reduced following the same standard procedure outlined in the ARAS web pages %see also http://www.astrosurf.com/buil/isis/guide\_lisa/tuto\_en.htm) 
and calibrated in flux through simultaneous photometry observations. %(http://www.astrosurf.com/buil/calibration2/absolute\_calibration.htm). 
The HEROS spectra, although pipeline processed according standard procedures, were not flux calibrated. 

The last sequence of spectra consists of two NOT/FIES (Telting et al. 2014, Frandsen \& Lindberg 1999) spectra taken about two years after outburst with the aim of observing the optically thin nebular phase of the object (see Table\,\ref{fieslog}). The data were reduced with the instrument pipeline at the telescope. The observing strategy envisioned for the instrument did not include background subtraction (there is no dedicated sky-fiber in mid-resolution mode), since it is expected to be insignificant. We verified this with a 900 sec sky exposure in which both the exposed and the masked part of the CCD showed the same count level, and with the science exposure whose inter order background roughly matched that of the sky exposure. We therefore ascribe only a few percent uncertainty to the flux calibration process in the absence of sky subtraction with FIES. We note that the largest systematic effect is introduced by the reddening correction (see Section 2.1).

Of the two FIES epochs, the May spectrum was contaminated by stray light (John Telting, private communication) that affected the background level redward of 6400 \AA. 
Therefore, any physical parameter derived in the following sections relies on the data taken during the July run, which happened after the installation of a new light-leak baffle around the spectrograph shutter. 
During the July run, however, we experienced a color loss between the first and following two exposures. Specifically, we found up to 30\% loss in the blue part of the spectrum (H$\gamma$, 4363\AA), and only $\sim$5\% at H$\alpha$ and 7065 \AA, in the line flux of the second and third exposure compared to the first. This happened because the spectrograph atmospheric dispersion corrector (ADC) is set at the beginning of the sequence and not updated during subsequent exposures. Since the target crossed the meridian after $\sim$2/3 of the first exposure, the first exposure should have suffered differential color losses as well. Thus, our derived density (see Section\,6) is only a lower limit even if we use only the first spectrum in the calculation.  

\begin{table}[]
    \caption{Epoch and instrument setup for the NOT (Nordic Optical Telescope)/FIES (FIber fed Echelle Spectrograph) observations. The choice of fiber bundle 3 med-red, corresponds to a resolving power R$\sim$46000. }\label{fieslog}
    \centering
    \begin{tabular}{lcccc}
    \hline\hline
UT date & age & UT start & exptime & setup  \\
        & (d) &  & (s) &  \\ 
      \hline
2019/05/28 & 707 & 02:49 & 2200$\times$3 & F3 med-res \\ %JD (mid run) = 2458631.642988
2019/07/21 & 762 & 23:19 & 2650$\times$3 & F3 med-res \\ %JD (mid run)= 2458686.533587
      \hline
    \end{tabular}
\end{table}

\begin{figure}[h]
\centering
\includegraphics[width=9cm]{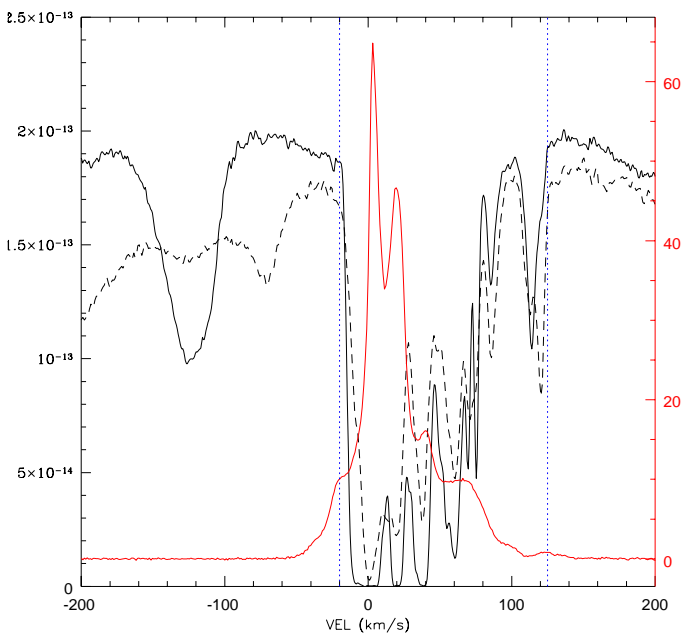}
\caption{Comparison of the temperature profile of the H~I 21 cm (red line) with the Na~I (black solid line) and Ca~II~K (black dashed line) interstellar absorption profiles. The two vertical blue dotted lines mark the limits for the column density calculation (see Section\,2.1 for details). Flux (erg/s/cm$^2$/\AA) and temperature (K) units are on the left and right axis, respectively.}
\label{lab21}%
\end{figure}

\subsection{Reddening correction of the spectra}

The high signal-to-noise and resolution of the UVES spectra allow us to use the interstellar absorptions to estimate the reddening. The spectra show multiple saturated Na~I~D and Ca~II~H\&K interstellar components, indicating that 17hx is significantly reddened. The saturation of the Na~I~D absorption lines prevent using the Munari and Zwitter (1997) EW(Na~I~D)-E(B-V) relation. The much weaker and resolved K~I absorptions are, however, blended with telluric absorptions so that using the analogous relation for the potassium resonance line provides only a lower limit: E(B-V)$>$0.4 mag. 

Fig.\ref{lab21} displays the antenna temperature profile of the neutral H from the LAB 21 cm maps (Kalberla et al. 2005) together with the interstellar absorption from Na~I and Ca~II in the local standard of rest (LSR) frame. The good match in the structures support using of the H~I column density to estimate the E(B-V) following Bohlin et al. (1978).  Specifically, integrating the H~I temperature in the velocity range [-20,+125] km/s (i.e. the velocity range spanned by the sodium and calcium absorptions), we derive E(B-V)=0.71$\pm$0.02 mag, where the uncertainty is purely statistical. This is the reddening we apply to all spectra in the following analysis. 
We note that the adopted E(B-V) agrees with the values derived by Munari et al.  (0.68 mag, 2017a) and by Kuin et al. (0.8$\pm$0.1 mag, 2017), within the errors. 

All figures in this work were created using the observed spectra (i.e. not corrected for reddening). We applied the reddening correction ahead of any physical parameter derivation (Section~\ref{numbers}). 

%--------------------------------------------------------------------
\begin{figure*}[ht]
\centering
\includegraphics[width=14cm,angle=270]{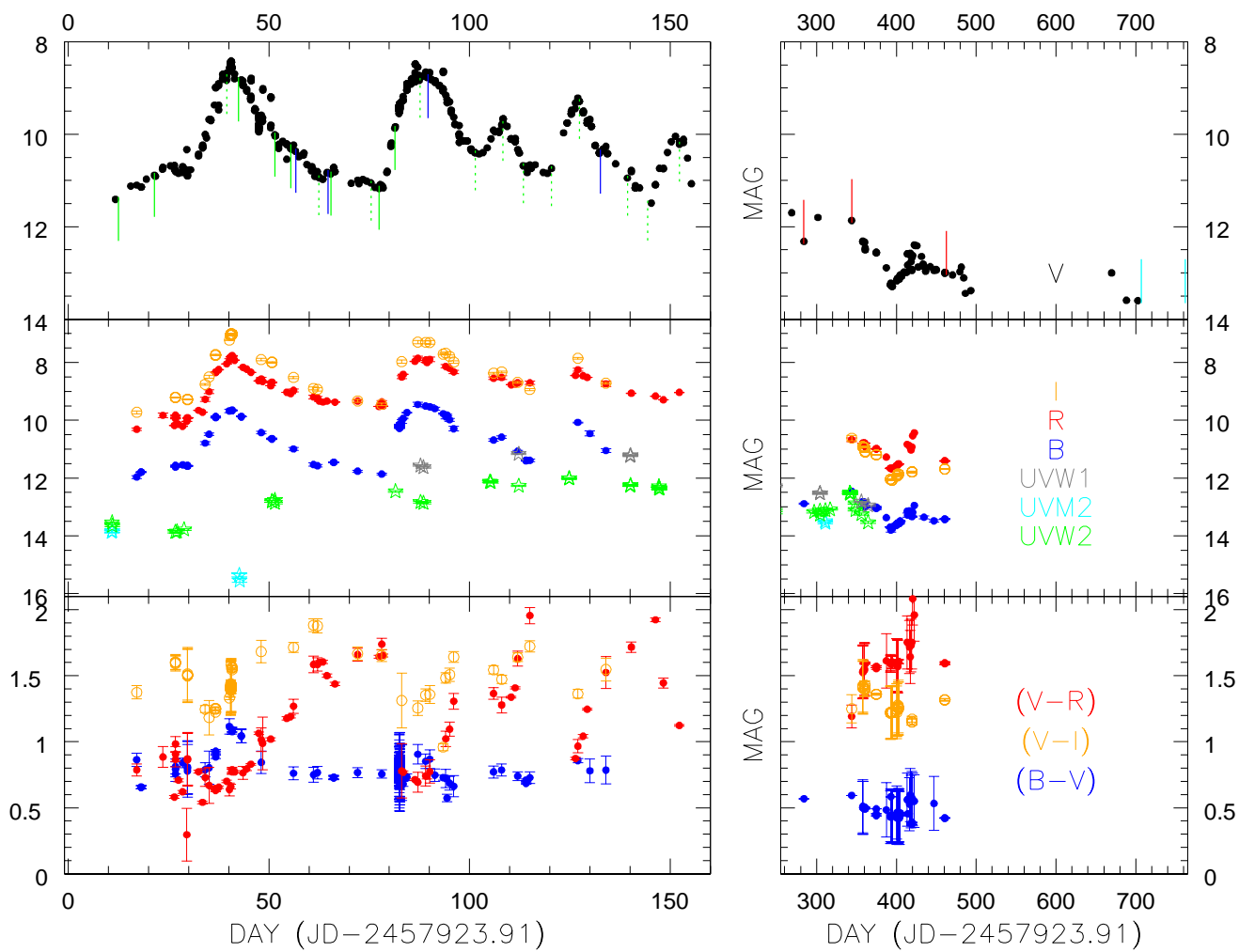} 
\caption{Top panels: the AAVSO V light curve together with the epoch of the UVES (blue lines), ARAS (green dotted and solid lines, representing respectively the high resolution un-calibrated and low resolution flux calibrated spectra), the LISA and HEROS observations (red lines) and the FIES spectra (cyan lines). Mid panel: the AAVSO B, R, and I light curves together with the {\it Swift} UVOT UVW1 ($\lambda_c$=2600 \AA), UVM2 ($\lambda_c$=2246 \AA), and UVW2 ($\lambda_c$1928 \AA). Bottom panel: the color light curve: (B-V), (V-R) and (V-I). Error bars $\geq$0.2 mag in the color light curves were arbitrarily set whenever either the V or the paired broad band photometry had not associated error bar. Colors were computed every time the V and broad band filter were taken within 3 hr observations.}
\label{lc}%
\end{figure*}

\begin{table}[]
    \caption{Swift UV photometry. Columns 2 and 3 have been rounded to the closest integer; 4 and 5 to the second decimal.  }\label{uvphot}
    \centering
    \begin{tabular}{lccccc}
    \hline\hline
MJD & age & exptime & mag & mag err & filter \\
  &  (d)  &    (s)  &     &          &      \\ 
      \hline
57934.272744 & 11 & 147 & 13.87 & 0.04 & UVM2 \\  
57934.339648 & 11 & 107 & 13.85 & 0.04 & UVM2 \\
57966.028409 & 43 & 129 & 15.37 & 0.06 & UVM2 \\
57966.096473 & 43 & 127 & 15.55 & 0.07 & UVM2 \\
58234.180839 & 311 & 187 & 13.50 & 0.03 & UVM2 \\
58234.374123 & 311 & 149 & 13.56 & 0.03 & UVM2 \\
58743.536150 & 820 & 228 & 15.55 & 0.05 & UVM2 \\
58746.780652 & 823 & 102 & 15.51 & 0.08	& UVM2 \\
\hline
57934.286262 & 11 & 93 & 13.54 & 0.03 & UVW2 \\ 
57934.356908 & 11 & 124 & 13.66 & 0.03 & UVW2 \\ 
57950.042452 & 27 & 103 & 13.87 & 0.03 & UVW2 \\ 
57950.507166 & 27 & 81 & 13.86 & 0.04 & UVW2 \\ 
57952.234154 & 29 & 97  & 13.78 & 0.03 & UVW2 \\ 
57974.204695 & 51 & 30 & 12.80 & 0.04 & UVW2 \\ 
57974.216188 & 51 & 82 & 12.81 & 0.03 & UVW2 \\ 
57974.865251 & 51 & 6 & 12.77 & 0.07 & UVW2 \\ 
57974.876617 & 51 & 80 & 12.86 & 0.03 & UVW2 \\ 
58004.979097 & 82 & 130 & 12.46 & 0.03 & UVW2 \\ 
58011.211962 & 88  & 92 & 12.83 & 0.03 & UVW2 \\ 
58011.945289 & 89 & 93 & 12.87 & 0.03 & UVW2 \\ 
58028.483596 & 105 & 74  & 12.12 & 0.03 & UVW2 \\ 
58028.695898 & 105 & 61 & 12.15 & 0.03 & UVW2 \\ 
58035.666692 & 112 & 114 & 12.27 & 0.03 & UVW2 \\ 
58048.215990 & 125 & 115 & 12.01 & 0.03 & UVW2 \\ 
58048.276655 & 125 & 74 & 12.02 & 0.03 & UVW2 \\ 
58063.498629 & 140 & 115 & 12.27 & 0.03 & UVW2 \\ 
58063.565482 & 140 & 84 & 12.24 & 0.03 & UVW2 \\ 
58070.597671 & 147 & 43 & 12.36 & 0.03 & UVW2 \\ 
58070.609255 & 147 & 80 & 12.29 & 0.03 & UVW2 \\ 
58070.717840 & 147 & 38 & 12.36 & 0.03 & UVW2 \\ 
58070.732031 & 147 & 104 & 12.38 & 0.03 & UVW2 \\ 
58171.161775 & 248 & 123 & 13.08 & 0.03 & UVW2 \\ 
58171.219816 & 248  & 56 & 13.20 & 0.04 & UVW2 \\ 
58220.450239 & 297  & 78 & 13.18 & 0.03 & UVW2 \\ 
58227.547373 & 304 & 93 & 13.13 & 0.03 & UVW2 \\ 
58227.612017 & 304 & 83 & 13.27 & 0.03 & UVW2 \\ 
58234.386238 & 311 & 97 & 13.12 & 0.03 & UVW2 \\ 
58240.500602 & 317 & 134 & 13.09 & 0.03 & UVW2 \\ 
58265.014546 & 342 & 125  & 12.52 & 0.03 & UVW2 \\ 
58265.465045 & 342 & 157 & 12.55 & 0.03 & UVW2 \\ 
58265.477300 & 342 & 81  & 12.51 & 0.03 & UVW2 \\ 
58272.050910 & 349 & 82 & 12.88 & 0.03 & UVW2 \\ 
58272.109245 & 349 & 82 & 13.11 & 0.03 & UVW2 \\ 
58279.560809 & 356 & 52 & 13.25 & 0.04 & UVW2 \\ 
58287.995966 & 365 & 96 & 13.53 & 0.03 & UVW2 \\ 
58743.532107 & 820 & 294 & 14.86 & 0.03 & UVW2 \\ 
58746.790646 & 823 & 56 & 14.84 & 0.06 & UVW2 \\ 
58747.653168 & 824 & 53   & 14.80 & 0.07 & UVW2 \\ 
\hline
58011.198225 & 88  & 183 & 11.57 & 0.02 & UVW1 \\ 
58011.931645 & 89  & 168 & 11.63 & 0.02 & UVW1 \\ 
58035.649854 & 112 & 259 & 11.17 & 0.02 & UVW1 \\ 
58063.480888 & 140 & 179 & 11.24 & 0.02 & UVW1 \\ 
58063.551767 & 140 & 171 & 11.21 & 0.02 & UVW1 \\ 
58171.143848 & 248 & 218 & 12.23 & 0.02 & UVW1 \\ 
58171.210724 & 248 & 183 & 12.28 & 0.02 & UVW1 \\ 
58227.534013 & 304 & 121 & 12.53 & 0.03 & UVW1 \\ 
58227.600152 & 304 & 93  & 12.53 & 0.03 & UVW1 \\ 
58279.550702 & 356 & 233 & 12.83 & 0.02 & UVW1 \\ 
58287.981763 & 365 & 149 & 12.95 & 0.03 & UVW1 \\ 
58743.527626 & 820 & 146 & 14.42 & 0.04 & UVW1 \\ 
58747.643648 & 824 & 135 & 14.27 & 0.04 & UVW1 \\ 
\hline
\end{tabular}
\end{table}

\section{Data Analysis: the light curves}
Fig.\ref{lc} shows the 17hx light curves in the filters B,V,R and I, obtained from the AAVSO database (Kafka 2019), together with the epochs of the spectroscopic observations listed in Tables\,1-4 (top and mid panels). The bottom panel of Fig.\ref{lc} displays the  color curves. 
The middle panel of Fig.\ref{lc} also displays the sparse UV photometry obtained by {\it Swift}. The UV photometry is also listed in Table~\ref{uvphot}, including data points that are outside the plotted time intervals.  %Although only two data points were obtained with the UVM2 filter ($\lambda_c$=2246 \AA), they show that the first optical maximum corresponds to a UVM2 minimum. 
%{\bf Although only two data points were obtained with the UVM2 filter ($\lambda_c$=2246 \AA), the difference between UVM2 and UVW2 ($\lambda_c$=1928 \AA) is 0.3 mag and the two data sets together show that the UV flux drops when the optical magnitude reach a maximum. }
%This is consistent with flux redistribution of the radiation coming from a central object and also implies that the optical maxima do not result from additional energy input. This behavior was first noted in nova Cyg 1992, using extensive UV spectrophotometry and optical photometry (Shore et al. 1994). 

The optical photometry strikingly shows that 17hx light variations are exceptional in both amplitude and timescale. 
%Some interesting features are immediately apparent from the optical photometry. First, 17hx shows light variations at maximum that are exceptional in both amplitude and timescale. 
Brightness variations of CNe during maximum rarely exceed 1-1.5 magnitudes, while their time scales never extend beyond a couple of weeks. Here the amplitudes are $>$2 mag on a timescale of 1-2 months. The amplitudes are greater in the V band. %Second, those variations are unlike the typical "jittering" CNe during their maximum (see, for example, the J-class of light curves identified by Strope et al. 2010). 

The UV photometry, instead, is possibly suggestive of flux redistribution since the UVM2 flux displayed a drop during the first $\sim$50 days, while the optical flux increased. Unfortunately, because of the limited number of UV observations we cannot be conclusive. As of today, flux redistribution has been convincingly shown only for nova Cyg 1992 (Shore et al. 1994).

\begin{figure*}[ht]
\centering
\includegraphics[width=14cm,angle=270]{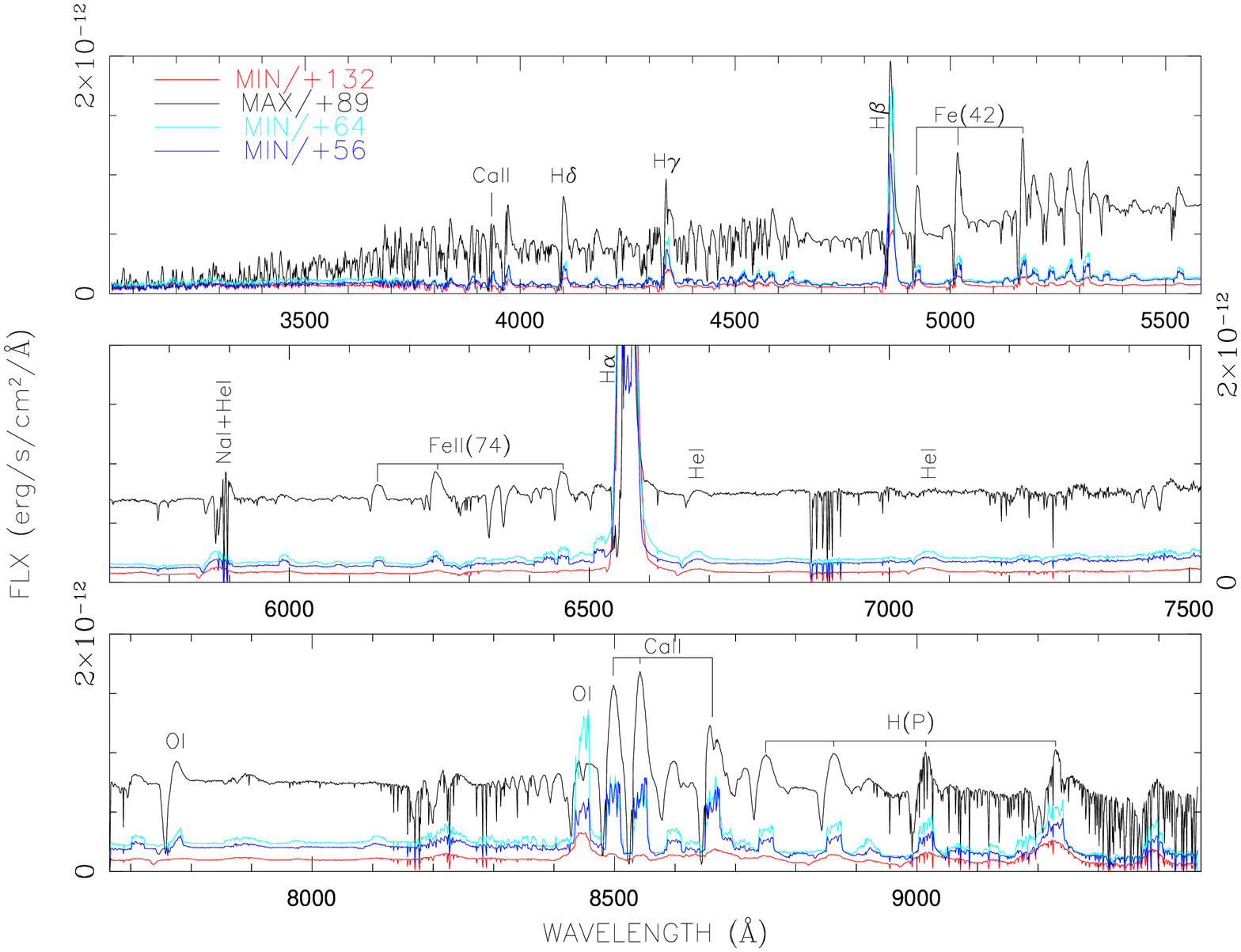}
\caption{The four high resolution spectra taken at VLT/UVES. The spectra have been split in three panels where there is the wavelength gap induced by the dichroic or the detector mosaic (red arm only). Photometric state and age since discovery (in days) are color coded in the upper panel. Note that H$\alpha$ is saturated in all UVES spectra.The spectra in the figure have not been corrected for reddening.}
\label{uvesall}%
\end{figure*}

The color light curves (Fig.\ref{lc}, bottom panel), show B-V, V-R and V-I colors that are in low amplitude anti-phase with respect to the broad band light curves; 17hx seems redder during the minima than the maxima. However, spectroscopy shows (see Section 4, Fig.\ref{uvesall} and \ref{spcseq}) that the continuum SED is redder at maximum than at minimum. The color variation and in particular the V-I and B-V color maxima are driven by the strong emission lines during the photometric minima. 
%We underline "seems" because, in fact, both the UV photometry mentioned above and the spectroscopy discussed in the next Sections suggest the opposite.  The V-R and V-I color light curves are explained by the larger intensity of the emission lines during the minima. The dominant H$\alpha$ and OI $\lambda$8446 emission lines fall on the maximum of the R and I band transmission-curve, respectively, contributing significantly to the overall integrated flux. In the bluer portion of the spectrum, instead, the dominant lines are H$\beta$ and H$\alpha$ which fall on the wings of the B and V filters transmission curve, implying no significant  contribution. 
Although less pronounced, similar behavior is not uncommon among classical novae (e.g. nova Cen 2013 and nova Sgr 2015b; see AAVSO light curves for the photometry and Mason et al. 2018 for nova Cen spectroscopy; nova Cen and nova Sgr 2015b spectra are publicly available in the ESO archive), and highlight the potentially misleading inferences from photometry alone. 

%--------------------------------------------------------------------

\begin{figure}
\centering
\includegraphics[width=9cm]{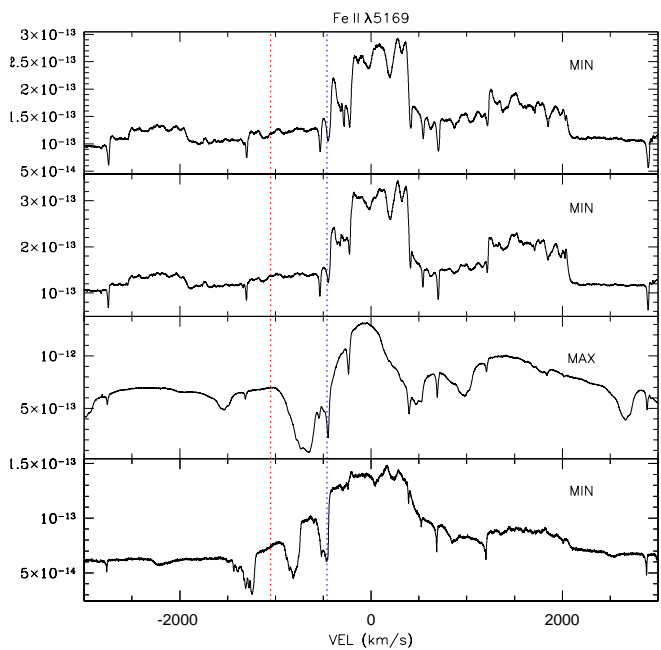}
\caption{Evolution of the Fe~II $\lambda$5169 (RMT 42) transition. Y axis units are erg/cm$^2$/s/\AA. The blue and red dotted lines mark the velocity of the persistent narrow absorption (see Section\,4.1.1) and of the He~I absorption in the first UVES spectrum, respectively (see text for details) for a comparison.  The photometric state (min, max) of each spectrum is indicated within each panel.}
\label{fe}%
\end{figure}

\begin{figure*}[h!]
\centering
\includegraphics[width=14cm,angle=270]{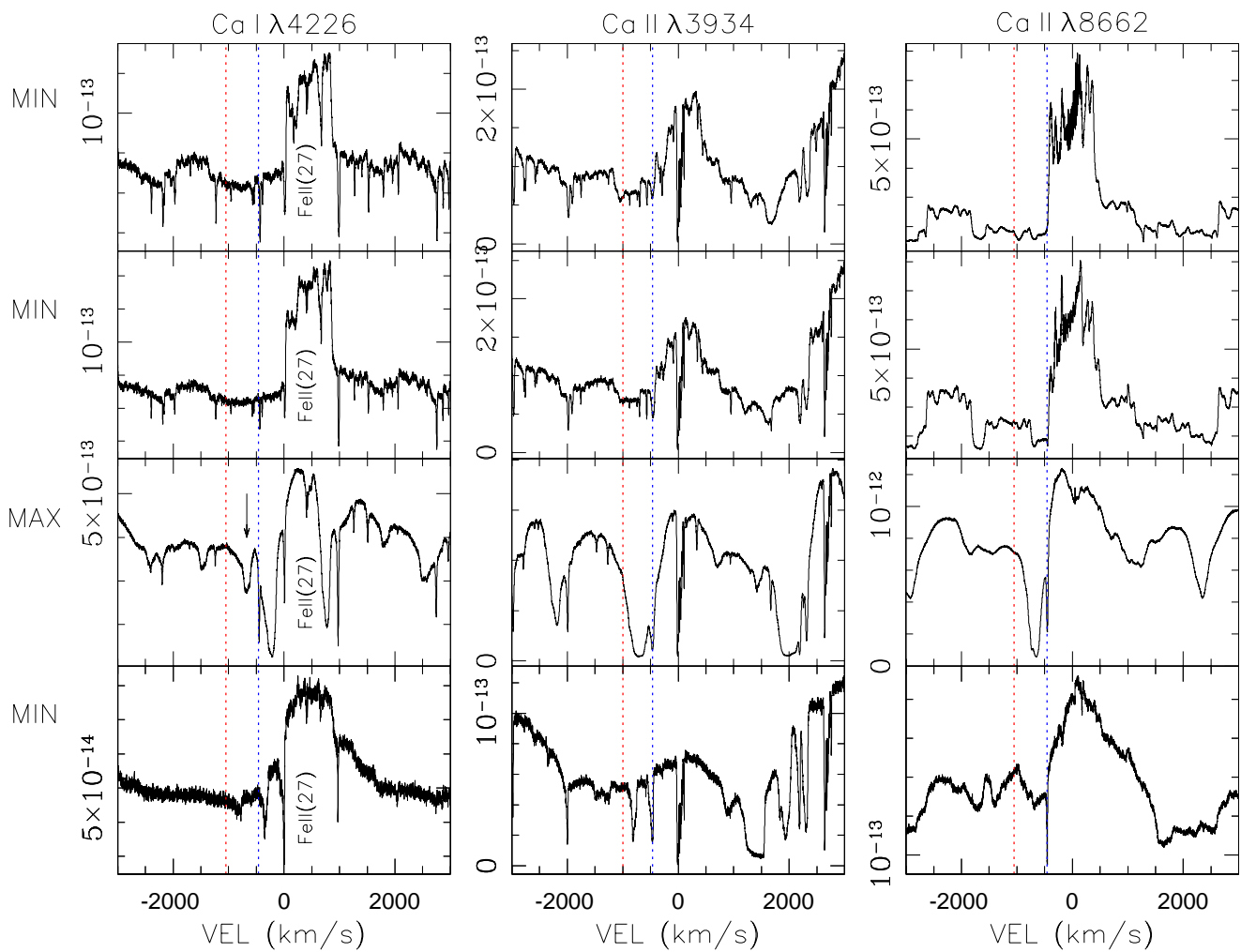}
\caption{Evolution of the Ca~I and Ca~II line profiles (see text for details). Note that the emission centered at about +500 km/s in the first column (Ca~I panels) is from Fe~II RMT 27. The arrow in the third panel from top indicates the second broad absorption from Ca~I $\lambda$4226. The Y axis units are erg/cm$^2$/s/\AA. The blue and red dotted lines mark the velocity of the persistent narrow absorption (see Section\,4.1.1) and of the He~I absorption in the first UVES spectrum, respectively. The photometric state (min, max) of the spectra is indicated on the left of each row.}
\label{ca}%
\end{figure*}

\begin{figure*}
\centering
\includegraphics[width=14cm,angle=270]{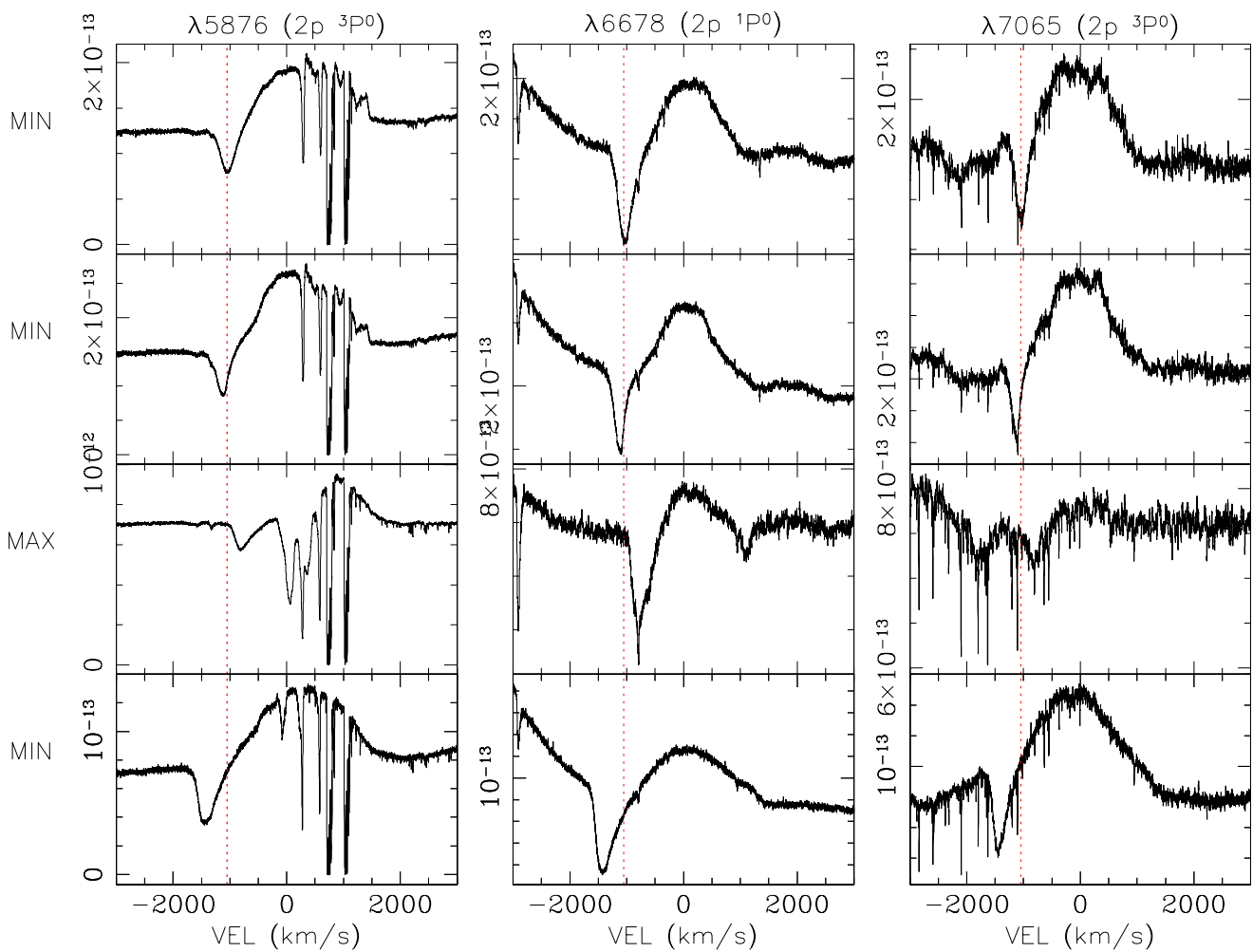}
\caption{Evolution of the He~I (triplet and singlet) transitions (see text for details). Y axis units are erg/cm$^2$/s/\AA. The red dotted line mark the velocity of the absorption component at the first epoch: $\sim$1050 km/s. The photometric state (min max) is indicated on the left of each row.}
\label{he}%
\end{figure*}

\begin{figure}
\centering
\includegraphics[width=9cm]{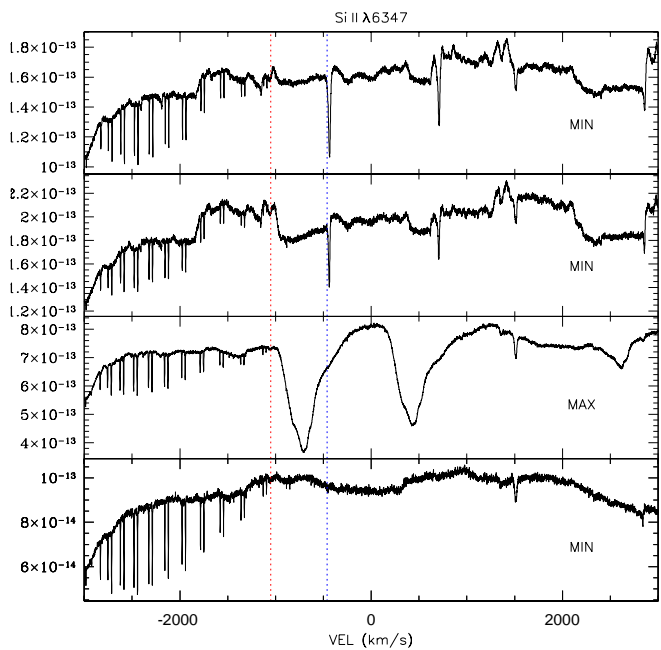}
\caption{Evolution of the Si~II doublet (RMT 2) transition (see text for details). Y axis units are erg/cm$^2$/s/\AA. The colored vertical lines are identical to those in the previous figures. The photometric state (min, max) of the  spectra is indicated within each panel.}
\label{si}%
\end{figure}

\begin{figure}
\centering
\includegraphics[width=9cm]{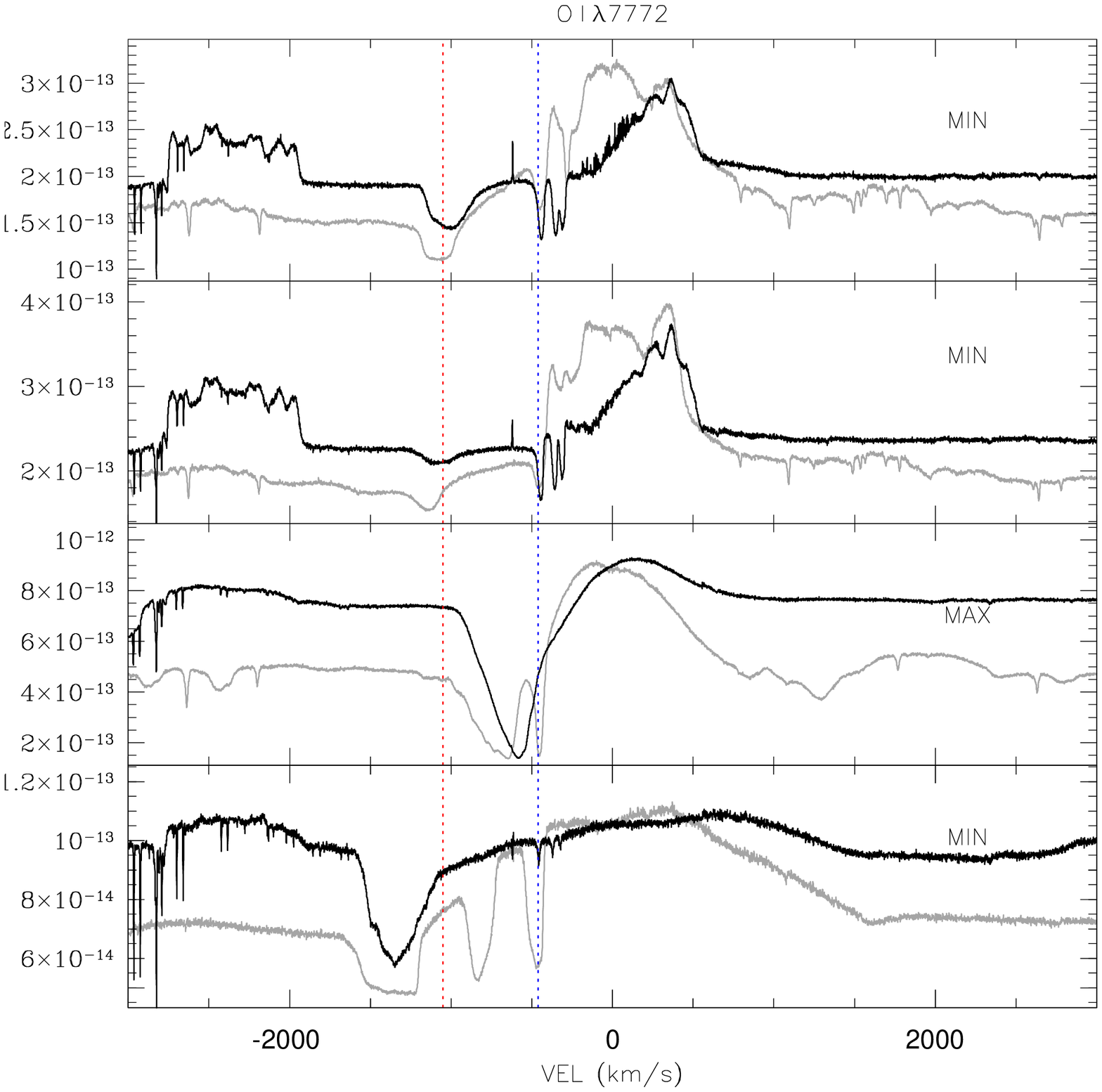} 
\includegraphics[width=9cm]{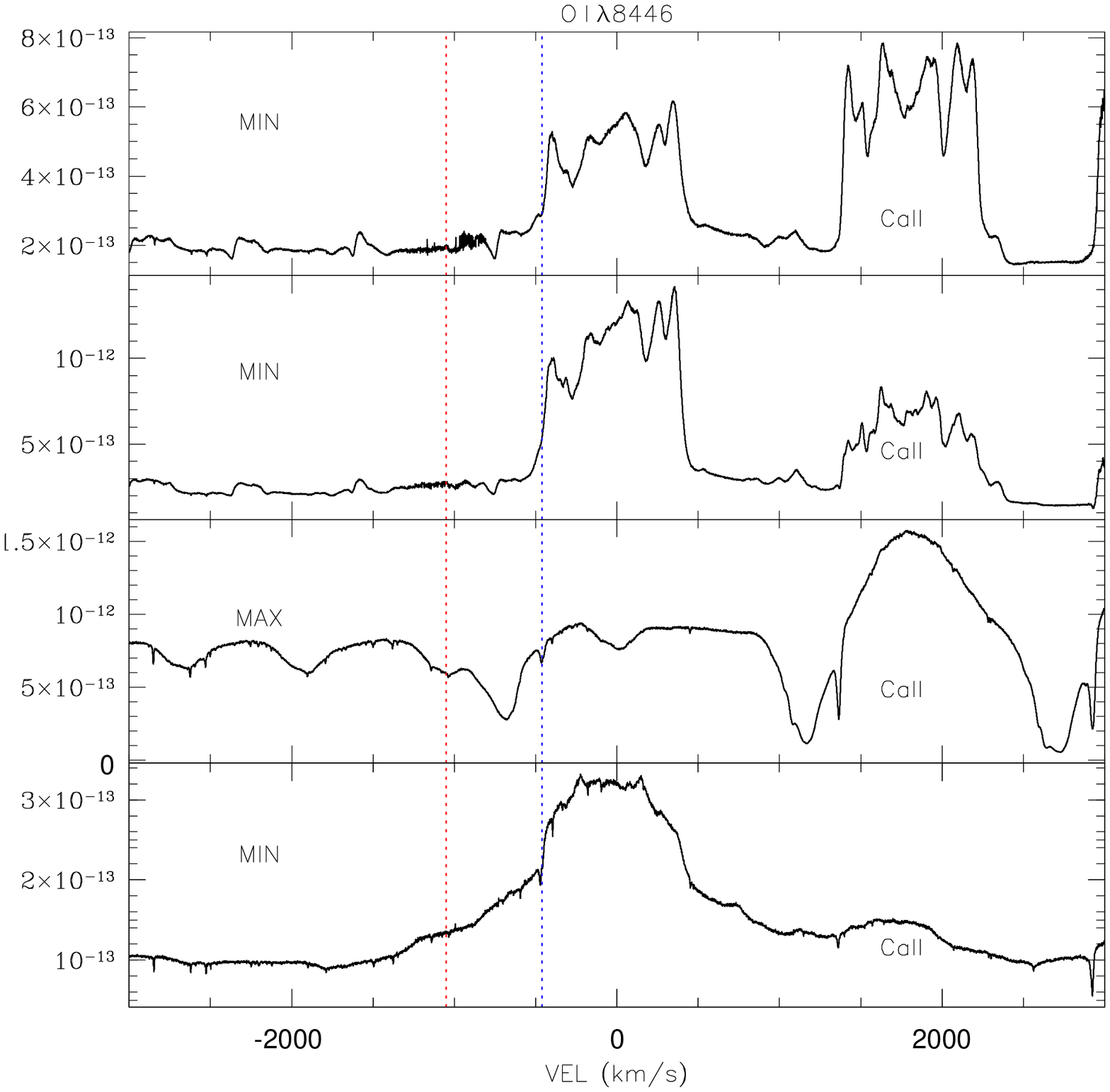}
\caption{Evolution of the O~I triplets (RMT 1, top, and 4, bottom). Note that triplet (1) is resolved in the narrow absorption at $\sim$-460 km/s. Together with the triplet (1) we overplot (gray line) the H$\delta$ profile for a comparison (see text for details; note that the H line has been arbitrarily offset in each subpanel; in the bottom subpanel it has also been scaled by a factor 0.5). Note that the blue line of the Ca~II triplet is included in the bottom panel and that it shows similar behavior to the O~I RMT 4. The Y axes units are erg/cm$^2$/s/\AA. The colored vertical dotted lines have the usual meaning. The photometric state (min, max) of each spectrum is indicated in each subpanel.}
\label{oi}%
\end{figure}

\section{Data Analysis: early spectroscopy, the optically thick stage}

\subsection{The high resolution UVES spectra}
The four UVES spectra cover the first decline and minimum after the first optical maximum, the second maximum, and the decline after the fourth maximum (see Fig.\ref{lc}). They therefore permit a comparison in great detail (R$\geq$60000) of spectral characteristics at the different photometric states providing important information about the gas kinematics and physics. 

At all epochs, the emitting gas is dominated by H~I (from the Balmer and the Paschen series, see Fig.\ref{uvesall}) and low ionization potential metal lines,  mainly Fe~II  (e.g. Revised Multiplet Table, RMT, 73, 40, 46, 55, 56, 49, 48, 41, 42, 35, 47, 37, 38, 27 and 28) but also Ca~I (RMT 2) and II (H\&K and the NIR triplet). %In particular, we detect numerous Fe~II multiplets (i.e. Moore multiplets 73, 40, 46, 55, 56, 49, 48, 41, 42, 35, 47, 37, 38, 27 and 28, from red to blue optical wavelengths), Ca~II H\&K and the Ca~II NIR triplet.
Non-metal transitions are also present, e.g. O~I (RMT 1 and 4), Si~II (RMT 2), and He~I. There are no forbidden transitions since, as we will show later, the densities at these stages are very high. 

Apart from that, the three minimum/decline spectra are similar to each other but very different from that at maximum. In particular, the minimum and decline spectra display a flat continuum, strong emissions with rectangular shape profile partially resolved into numerous structures (i.e. "castellated" tops), and weak or absent absorption components. Conversely, the maximum spectrum displays a redder continuum, weak emission components with a smooth profile and strong P Cyg-like absorptions (see Figs. \ref{uvesall}, \ref{fe}, and \ref{ca}). The difference in the emission line profiles suggests that, during minima, only a small number of structures is contributing to the emission component; while, at maximum, many more structures contribute to the emission creating a blended profile.

%remarkably different from that at maximum and very similar to each other. In particular, they display similar flat continuum and continuum levels and similar line profiles, especially for the metal lines\footnote{The H line profiles, although similar, reflect the increasing volume of scattering gas and, therefore, together with the minimum-maximum alternation they also display a secular trend (see Fig.\ref{cycle}).}, while the maximum spectrum shows a redder continuum and significantly different line profiles (see Fig.\ref{uvesall}). The intensity of the emission component is also strikingly different at minima and maxima, being stronger above the continuum (especially in the H Balmer transitions) in the minima spectra. 

Most important, the ionization degree of the emitting gas changes with the photometric oscillations and is higher at minima than at maximum.  This is best illustrated by the changes in the He~I lines (see Fig.\ref{he}).
%We will go over these difference in detail in the following paragraphs.
%
%First we consider the line profiles. Fig.\ref{fe} and \ref{ca} (but see also Fig.\ref{uvesall}) show the evolution of the metal line profiles with time. During the minimum/decline states the line profiles are clearly rectangular emission (FWHM$\simeq$FWZI$\sim$1000 km/s) with numerous structures (i.e. castellated tops). They consist of a smoother P Cyg profile with a pronounced absorption component and a weaker bell-shaped emission component, during maximum. 
%The profiles observed at maximum suggest a large number of emitting structures contributing to the whole emission component with their blending producing the smoother profile. Instead, at minimum only a smaller number of partially resolved structures contribute to the emission component.  
%Note that both the minimum and the maximum line profiles also show narrow (FWHM$\sim$20-30 km/s) absorption at $\sim$-460 km/s that appears to be stationary (we will come back to this later in this section).
%
%Second, the most obvious indication of increased ionization at minima comes from the He~I lines (see Fig.\ref{he}).
%They are characterized by strong broad absorptions at a relatively low velocity ($\sim$-750 km/s, FWHM$\sim$300 km/s) and weak to absent emission (this is particularly true in the triplet transitions) in the maximum spectrum, in contrast to a strong and broad emission component (FWHM$\geq$1000 km/s) with weak absorption at relatively high velocities ($\geq$-1000 km/s) during the minima.  
The weakness of the emission component during the maximum indicates reduced He$^+$ recombination, yielding insufficient emission measure for detection. The velocity displacement of the absorption, instead, indicates that during maximum the absorption (of high energy photons) is confined to the low velocity regions of the gas. In other words, larger volumes and higher velocities are involved in the He$^+$ recombination during minima.  

The Ca transitions confirm this trend (see Fig.\ref{ca}). 
The strengthening of the narrow absorption at $\sim$-460 km/s, together with the appearance of a much broad and deep absorption component at $\sim$-660 km/s in both Ca~I and Ca~II transitions indicates more neutral Ca and Ca$^+$ 
% i.e. more recombining Ca$^{++}$
at maximum than at minimum. During the photometric minima the Ca is largely twice ionized, given that the Ca$^+$ ionization potential (IP) is $\lesssim$12 eV, i.e. less than that of H$^0$, Fe$^+$ and much less than that of He $^0$. 
%Ca~I $\lambda$4226 (resonant) is present in all the spectra with only the narrow absorption at $\sim$-460 km/s. The absorption is relatively weak at the minima and an order of magnitude stronger at maximum. The Ca~II H\&K and the NIR triplet emission components display only the narrow absorption at $\sim$-460 km/s during the minima and a strong, broad P Cyg absorption during the maximum. This suggests that there is more Ca$^+$ at maximum than at minimum light, when the Ca is largely twice ionized (the Ca$^+$ ionization potential, IP, is $<$12 eV, i.e. less than that of H$^0$, Fe$^+$ and much less than that of He $^0$).  

Neutral silicon, an intermediate ionization potential element (IP$\sim$8 eV), displays  analogous behavior to the Fe~II transitions (see Fig.\ref{si}). Like the optical iron multiplets, the Si~II doublet is a transition between two excited levels, specifically 4s and 4p. The lower level 4s is connected to the ground state by resonance transitions ($\lambda$1533,1526). To produce $\lambda$6347,6371 emissions and absorptions, the resonance UV transition must be optically thick with low collisional de-excitation of the 4p level. This suggests an electron density n$_e\simeq$7$\times$10$^{14}$ cm$^{-3}$.  Again, the strong broad absorption visible at maximum suggests that at this time there is more Si$^+$ than at minima when it is likely twice ionized.  
%Similarly to the Fe~II transitions, the Si~II doublet shows the narrow absorption at $\sim$-460 km/s and structured emission component during the minima, which evolve into a very broad absorption displaced toward higher velocities ($\sim$-750 km/s) P Cyg profile at the maximum. There are more (blending) structures contributing to the emission component at maximum while, along the line of sight, there is little or no excited Si$^+$ during the minima (it is likely twice ionized), but there is a lot in the intermediate velocity range, during the maximum. 
The Si~II disappears in the last UVES spectrum, suggesting that the gas has diluted to the point that the UV doublet $\lambda$1553,1526 is no longer optically thick and thus incapable of overpopulating the 4s level. We exclude that the Si$^+$ further ionized into Si$^{2+}$ since the IP for Si$^+$ is about the same as Fe$^+$ ($\sim$16 eV) and Fe~II transitions are still present in the last UVES spectrum. 

Finally, O~I is detected only in the RMT triplets (1) and (4). They are interesting for their remarkably different profiles. 
%The O~I transitions are also noteworthy. We detect only the Moore triplets (1) and (4), which display remarkably different profiles. 
The RMT triplet (1) ($\lambda$7772,7774,7775) forms mainly via recombination cascade of O$^+$. It can display absorption components because its lower level leads to the ground state through an intercombination transition. In contrast, RMT triplet (4) (centered around $\lambda$8446), is mainly produced by optically thick H Ly$\beta$ transition that pumps the upper level of the OI $\lambda$1025 and, by cascade, populates the upper level of the $\lambda$8446 triplet (Keenan and Hynek 1950) in combination with an optically thick O~I $\lambda$1302 (which is responsible of the overpopulation of the lower level $^3$S of the triplet and explains the occasionally strong absorption component). Neutral oxygen has nearly the same ionization potential energy as H ($\sim$13 eV) and the $\sim\lambda$7774 triplet has a similar profile to the Balmer lines (see Fig.\ref{oi} where the triplet is plotted together with the H$\delta$ transitions, top panel). %The triplet shows the same absorption in the velocity range [-1200,-1000] km/s and the narrow absorption at $\sim$-460 km/s as the hydrogen lines, while the intermediate velocity emission structure at $\sim$-300 and +400 km/s are missing or much weaker. 
The hydrogen and oxygen recombine together with just small local differences. %Their profiles are also similar at maximum when the P Cyg absorption deepens and is confined to the smaller velocity range [-1000,-500]. 
The $\lambda$8446 triplet, in contrast, shows only emission whose profile matches the Fe~II and Si~II transitions at minima,  but displays almost pure absorption  at maximum.
%(with velocity  in the range [-750,-500] km/s) at maximum. 
Like Fe~II and Si~II, the $\lambda$8446 triplet results from UV pumping that is favored by the high UV opacity of the "iron curtain". 

In summary, at minimum the ionization state of the gas increases displaying stronger, structured emission profiles and higher velocity absorptions in the high IP energy transitions, but null or weak absorption in the low IP energy transitions. In contrast, at maximum the gas recombines, the average emission measure within each transition diminishes, and the photoexcitation is confined to the low velocity range (the lower deep absorptions of the maximum P Cyg profiles).  

\subsubsection{The narrow absorption at around -460 km/s}

All the UVES spectra display a narrow absorption at $\sim$-460 km/s that is present in all transitions (H~I, Fe~II, Ca~I and Ca~II, Si~II, K~I, O~I) except those of He~I (see the blue dotted line in Figs.\ref{fe}, \ref{ca}, \ref{si} and \ref{oi}). It persists also when other absorption components disappear.  Although this could be consistent with circumstellar gas, we exclude this explanation for the following reasons. Absorptions from excited and/or metastable levels (as in the case of Fe~II and Ca~II NIR) would imply unusually large column densities.  Similarly, the detection of the narrow absorption in the O~I $\lambda$7774 and H Balmer transitions would require that the central object is sufficiently powerful to ionize the circumstellar environment that is more than 114 AU = 2.4$\times$10$^4$ R$_\sun$ away from the transient\footnote{Assuming an expansion velocity of $\sim$1500 km/s as observed in the He~I lines and derived from our bicone model (see Fig.\ref{he} and Section 6)}, since it is dynamically undisturbed at day 132. 
Figs.\ref{fe} to \ref{ca} and EW measurements show that the narrow absorption is stronger at maximum and weaker at minimum; i.e. it varies responding to the radiation field in phase with the rest of the ejecta. This requires similar physical conditions in the narrow absorption region and the ejecta.  
Hence, the narrow component originates in the ejecta and its persistence informs about the gas kinematics,  specifically, that it is in ballistic expansion ($v\sim r$) as typical for explosive processes. The emitting gas is not a wind and is not a pulsating pseudo-photosphere but an ejecta similar to CNe.  

\begin{figure*}
\centering
\includegraphics[width=14cm,angle=270]{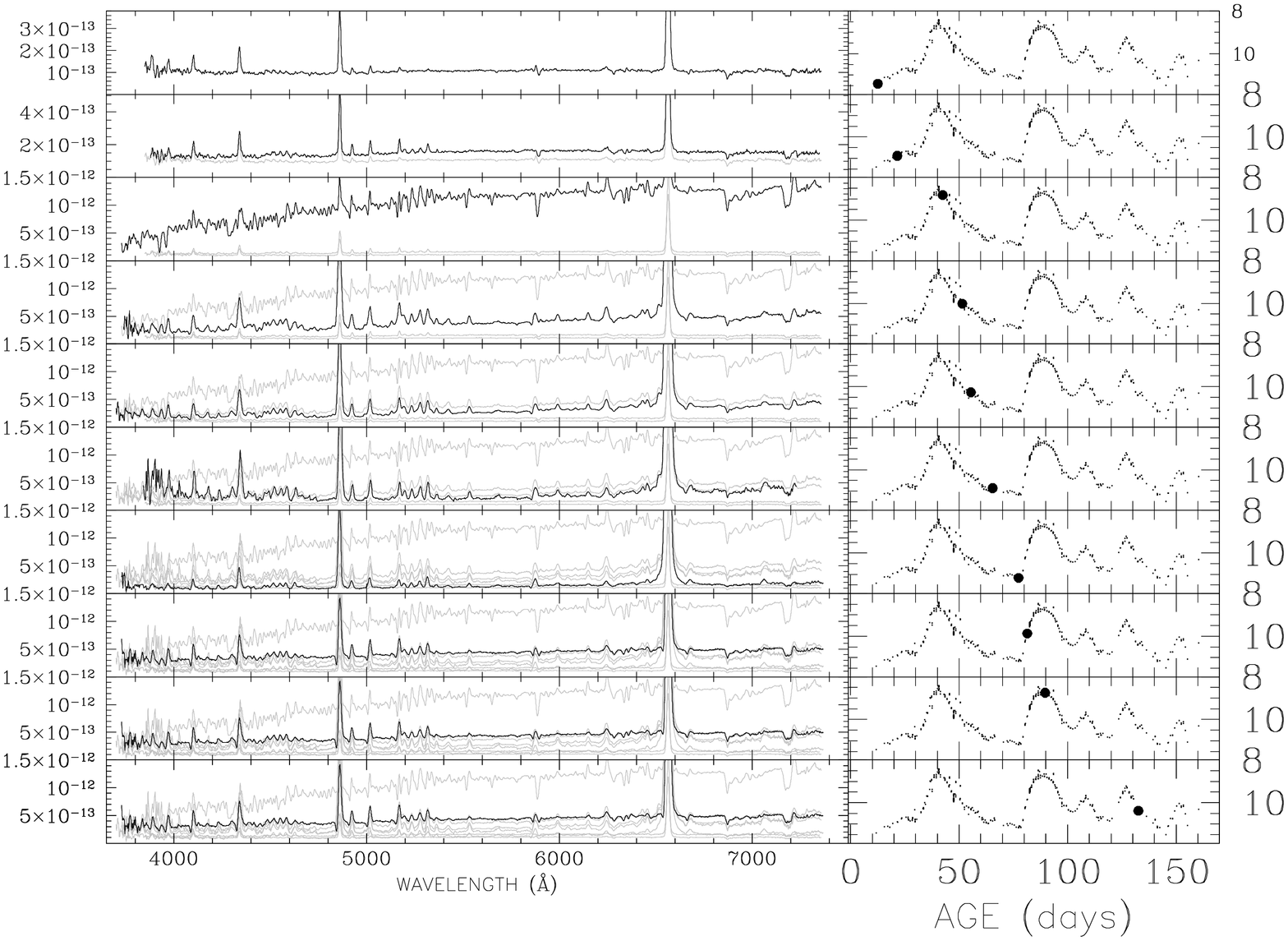}
\caption{The sequence of low resolution flux calibrated ARAS spectra, together with their position in the V-band light curve. Y axis units are erg/cm$^2$/s/\AA \, and mag on the left and right side, respectively. Each panel plots in gray the previous spectra for an easier comparison. }
\label{spcseq}%
\end{figure*}

\subsection{The low and medium resolution ARAS spectroscopic sequences}
The ARAS spectra listed in Table~\ref{araslog} support and extend our finding from the UVES spectra. Fig.\ref{spcseq} shows the evolution of the SED, uncorrected for reddening, and of the emission lines across the first photometric cycle. 17hx displays a flat continuum and strong emission lines with weak or no (i.e. not resolved) P Cyg-like absorptions  during the minima and the declining states. Instead, it has a redder continuum and strong P Cyg during maxima and the rising state.

\begin{figure*}
\centering
\includegraphics[width=17.6cm]{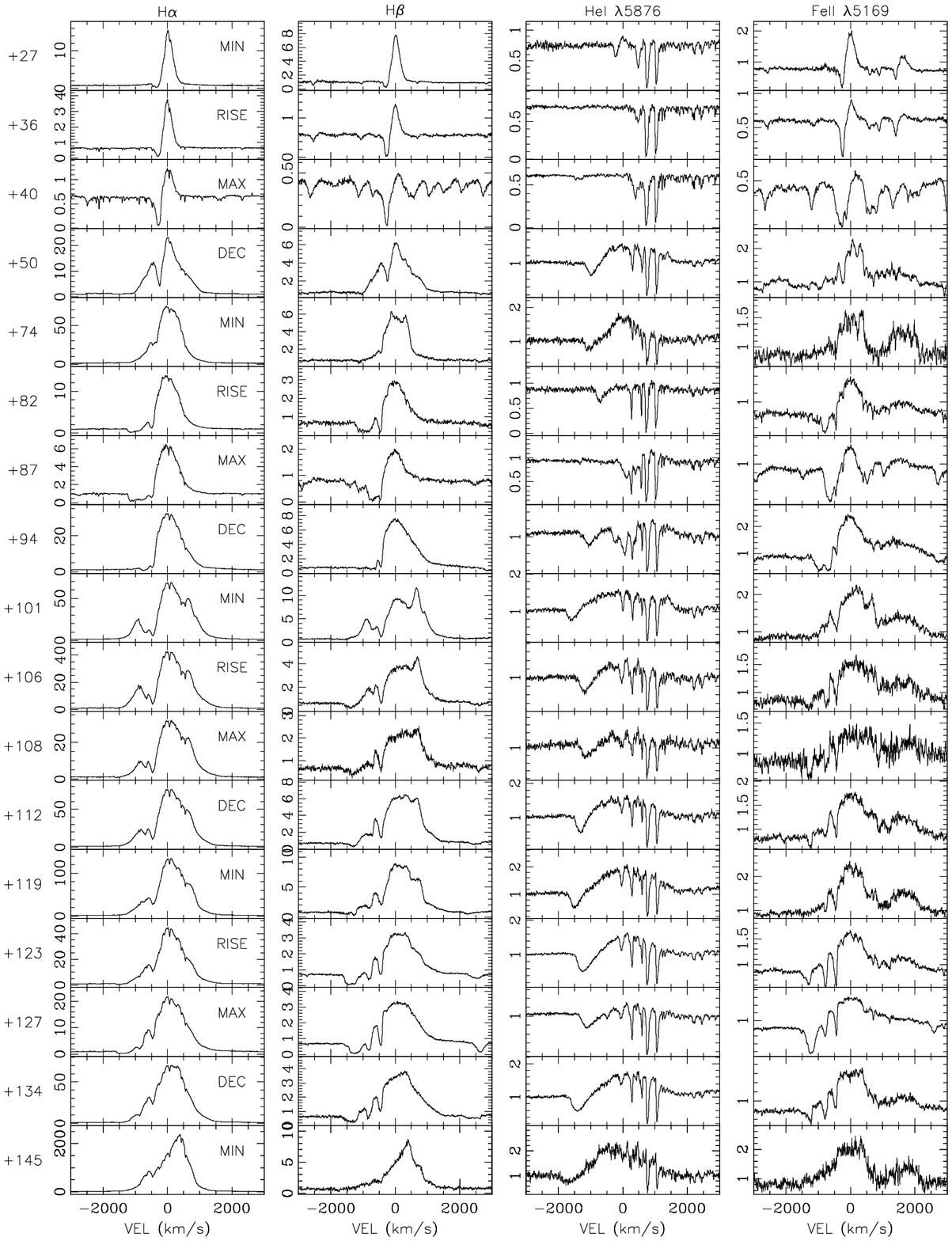}
\caption{The evolution across four min-max cycles of the H, He~I and Fe~II lines, representing respectively intermediate, high and low ionization potential energy atoms. The exact photometric state of each spectrum is indicated in the first column panels; the age (in days) is on the left. Note that the He~I transition is contaminated by Na~I~D both of interstellar and ejecta origin. The Na~I~D transition (emission and absorption) originating in the ejecta is particularly evident in the spectra taken between day 82 and 101.}
\label{cycle}%
\end{figure*}

The similarity of the line profiles at minimum and decline or maximum and rising phases is more evident in Fig.\ref{cycle} which displays the evolution of H$\alpha$, H$\beta$, He~I~$\lambda$5876 (blending at any time with Na~I~D both of ejecta and interstellar origin), and Fe~II~$\lambda$5169. The spectra are normalized to the continuum. %The time sequence runs from top to bottom for each transition with the age of each row indicated on the left side of the figure. The corresponding light curve state (i.e. minimum, maximum, rise or decline) is shown in the H$\alpha$ panel and applies to all the panels in the same row. 
What emerges from the figure is: 
\begin{enumerate}
    \item the emission line component in each transition at minimum is always at least an order of magnitude stronger than at maximum;
    %\item the H$\alpha$ emission component is exceptionally strong, at least an order magnitude stronger than H$\beta$ and two order of magnitude stronger than the other transitions;%\footnote{Note that H$\alpha$ was saturated in all UVES exposures.};
    \item the H$\alpha$ and H$\beta$ profiles are remarkably different (especially from day +74 on), indicating that the lines are very optically thick; 
    \item within each cycle, the line profiles change from a strong emission flanked by a weak P Cyg absorption to a weak emission with a strong P Cyg absorption. However, on subsequent cycles the emission component both strengthens and broadens; 
    \item the emission profiles differ systematically in structure depending on the photometric state, showing rectangular and castellated forms, especially in the low ionization potential energy transitions, during the minima and smoother and featureless profiles at maxima. 
    \item high ionization potential energy species (e.g. He$^+$ which recombining produces He~I lines) display no emission during maxima and detectable emission during minima. In addition, their absorption component (which is always single and shallow) moves outward and inward at minima and maxima respectively. 
\end{enumerate}

Both Fig.\ref{spcseq} and Fig.\ref{cycle} show that the degree of ionization of the emitting gas increases at minima with respect to maxima while the continuum becomes bluer. 
%--------------------------------------------------------------------
\begin{figure*}
\centering
\includegraphics[width=14cm,angle=270]{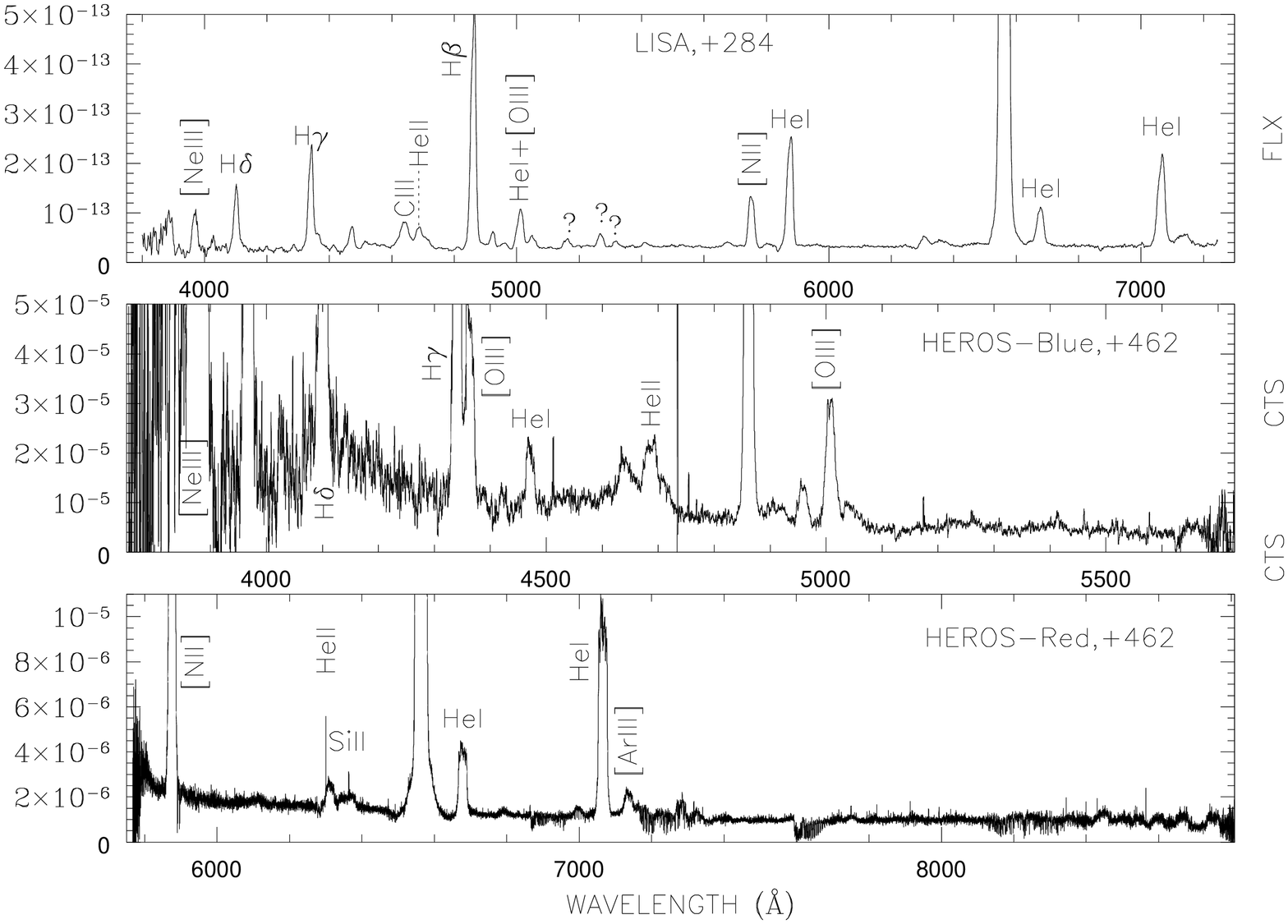}
\caption{The LISA (top panel) and HEROS (mid and bottom panel) spectra of 17hx $\sim$1 year after outburst/discovery. The LISA spectrum is in unit of erg/cm$^2$/s/\AA \ and uncorrected for reddening, while the HEROS spectrum is in arbitrary units. The spectrograph and the age of the spectrum are indicated in each panel.}
\label{1yr}%
\end{figure*}

\section{Data Analysis: late low and medium resolution spectroscopy, the transition stage}
The two LISA low resolution spectra taken during March and May 2018, 284 and 344 days after outburst, are similar to each other and show a highly opaque emitting region (see Fig.\ref{1yr} top panel). H$\alpha$ is one order of magnitude stronger than H$\beta$ in the observed spectrum (10$^{-10}$ vs 9$\times$10$^{-12}$ erg/cm$^2$/s/\AA, respectively), or a factor 5.3 stronger when dereddened. The strongest emission lines after the Balmer series are He~I, especially the triplets (e.g. $\lambda$5875 and 7065 \AA); while %whose intensity is about half that of H$\beta$). 
He~II $\lambda$4685 and the Bowen blend at 4640 \AA \, are weakly present. Weak forbidden transitions, such as [Ne~III] $\lambda$3968, [N~II] $\lambda$5755 and [O~III] $\lambda$5007, 4959 and 4363, are also present and evince the thinning of the gas. The [O~III] emissions are particularly weak and the $\lambda$5007 is blended with the strong He~I $\lambda$5015 producing a line centered at 5011 \AA. Even neglecting the He~I contribution to such a blend, the electron density resulting from the [O~III] line ratio (e.g. Osterbrock 1989, adopting T$_e\sim$10000 K) is $>$10$^7$ cm$^{-3}$, i.e. much greater than the critical value ($\sim$7$\times$10$^5$ cm$^{-3}$ for the [O III] RMT 2 and a few $\times$10$^3$ cm$^{-3}$ for the RMT 1; Osterbrock 1989). 
%The appearance of weak forbidden transitions demonstrates the thinning of the gas. In particular, we detect [Ne~III] $\lambda$3968 ($\sim$30\% H$\beta$ intensity in the dereddened spectrum), [N~II] $\lambda$5755 ($\sim$16\% H$\beta$ intensity) and very weak [O~III] $\lambda$4959 and 4363. The emission centered at 5011 \AA \ is a blend of [O~III] $\lambda$5007 and He~I $\lambda$5015. The relative intensity of the three [O~III] transitions  suggests that, even neglecting the significant contribution of He~I to the 5007 emission, the electron density is $>$10$^7$ cm$^{-3}$, i.e. much greater than the critical value. 
Other weak emission lines are  [O~I]$\lambda$6300,6364 and possibly Si~II $\lambda$6347. Weak unidentified emission lines are detected at 5159 \AA, 5267 \AA, and 5315 \AA. These cannot be coronal emission from, e.g., [Fe~VI], [Fe~V], [Fe~VII] respectively, since they are absent in the following HEROS spectra. They cannot be Fe~II, since the stronger multiplets (e.g. multiplet 42, 48, 49) should also be present; they are not, nor are there any detectable [Fe~II] transitions.

The higher resolution HEROS spectrum (day 462, Aug 20 2018; see Fig.\ref{1yr} bottom panels), although not flux calibrated, can be used to confirm the line identification of the LISA spectra and that the unidentified lines have disappeared. %In the few months between the LISA and HEROS observations, the relative intensity of the He~I lines has increased, while the three unidentified lines are gone. 
%The relative intensity of the [O~III] lines still indicates a high electron density and significant collision damping, although their intensity has started increasing relative to the other transitions. 
The higher resolution spectrum also reveals some differences in the line profiles. In particular, the forbidden transitions display relatively stronger wings (up to $\sim\pm$1200 km/s), indicating that they arise in the high velocity, lower density peripheral regions of the ejecta. 
%--------------------------------------------------------------------
\section{Data Analysis: late high resolution spectroscopy, the optically thin stage}

\begin{figure*}
\centering
\includegraphics[width=14cm,angle=270]{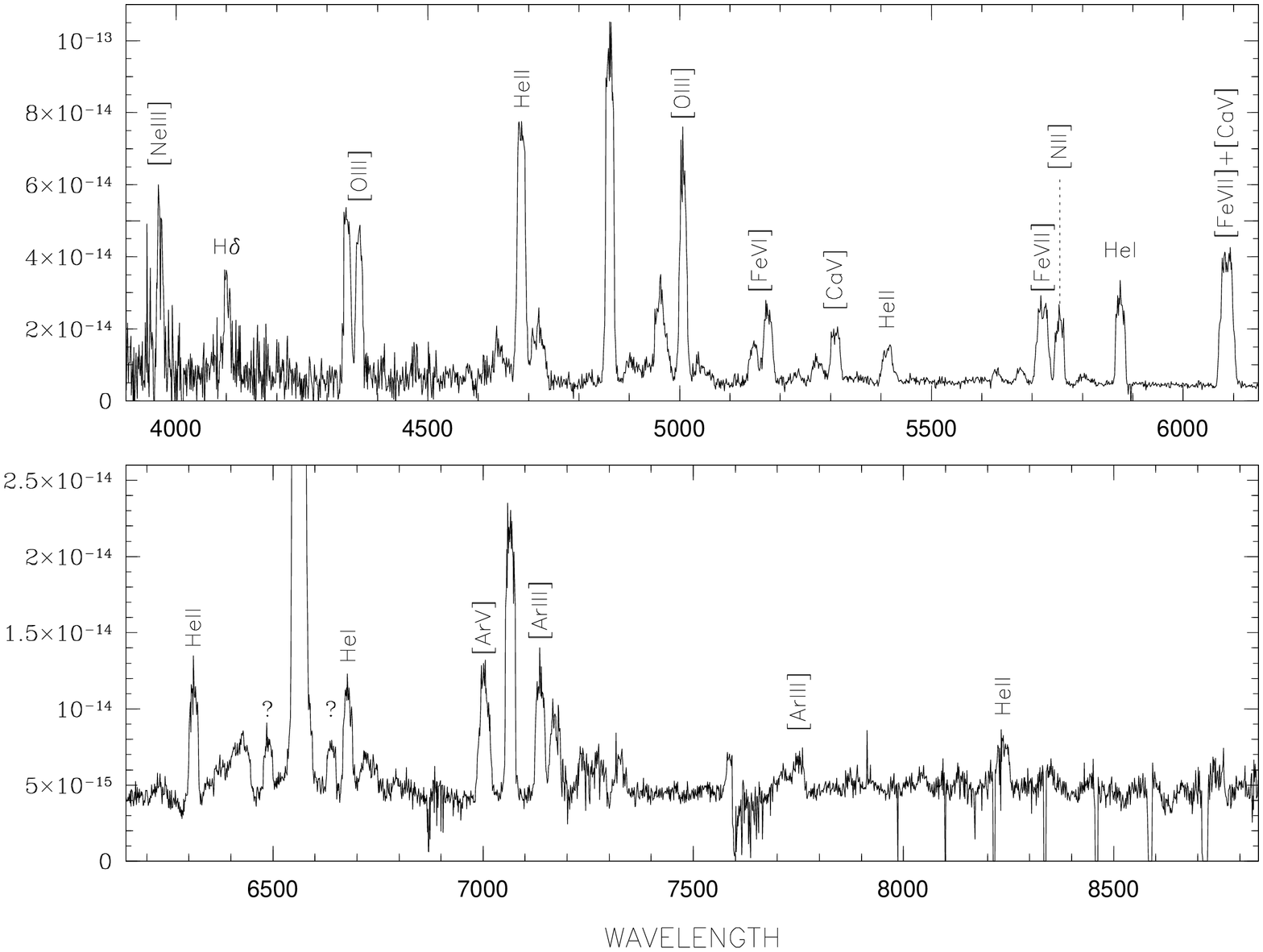}
\caption{The FIES spectrum of 17hx 762 days after discovery (uncorrected for reddening). The Y-axis units are  erg/cm$^2$/s/\AA}
\label{fies}%
\end{figure*}

The two FIES spectra, taken about 2 years after outburst, are separated by only $\sim$2 months. The spectra are quite similar and display the same species and transitions, in line with the slow evolution of 17hx. We show only the July 2019 spectrum in Fig.\ref{fies}, since it was not contaminated by scattered light (see Section 2). 
In both spectra, the strongest emission lines are still those of the H Balmer series. The forbidden transitions, however, have increased in relative intensity with respect to the 2018 observations, especially the  $\lambda$3869 of the [Ne~III](1) doublet that is among the strongest lines in the spectrum. The He~II $\lambda$4686 emission line is also one of the strongest lines (see Fig.\ref{fies}).
%now stronger than H$\beta$ (in the reddening corrected spectrum). The He~II $\lambda$4686 emission line rivals H$\beta$ in strength (see Fig.\ref{fies}). 
The [O~III] lines are present but weak. %, the $\lambda$5007 line intensity is only about half that of H$\beta$. 
We also identify numerous weaker emissions from He~I (triplet and singlet) and highly ionized metals: [Fe~VI] $\lambda$5177, 5147, 5236 and 5678; [Fe~VII] $\lambda$6086, 5276 and 5720; [Ca~V] $\lambda$5309 and 6086 (blending with [Fe~VII]). [Fe~X] $\lambda$6373 may be weakly present in a blend with other emissions of difficult identification. [N~II] $\lambda$5755 is present but weaker than He~I $\lambda$5876 and [Fe~VII] $\lambda$5720, while the fractional contribution of [N~II] $\lambda$6584,6548 to H$\alpha$ is insignificant. Other forbidden transitions are from [Ar~III] $\lambda$7136, 7751, [Ar~IV]$\lambda$7169 and [Ar~V] $\lambda$7006. There are a few lines which we are unable to identify, in particular, $\lambda$6638 and 6487, although are well isolated and not in a blend. We detect no C~II or C~III transitions and only very weak C~IV $\lambda$5801,5812. 

%The two FIES spectra differ only in the relative strength of some lines. Specifically, the H Balmer lines H$\beta$ and H$\alpha$ have slightly decreased in intensity from May to July. The He~I emission lines also decreased and at a somewhat larger pace, together with [N~II] $\lambda$5755. Instead, He~II, [Fe~VI], [Fe~VII] and [O~III] remained constant. The structures in the line profiles remain unchanged in the two epochs. 
The two FIES spectra confirm that 17hx entered an optically thin stage since the line profiles remain unchanged in the two epochs (see Fig.\ref{profile}). The ejecta however are not yet in nebular conditions how we will show in Section~\ref{numbers}.

\subsection{Central source spectral energy distribution constraints}

We can infer a few important things from the comparison of the 2019 FIES spectra with those from 2018, and the inter-comparison of the FIES spectra. 
The strengthening of the He~II emission together with the appearance of the coronal lines in the later spectra  indicates that the ionization degree has increased and that the ejecta is exposed to EUV/x-ray photons. The strengthening of the [O~III] emission lines, instead, indicates that the density has decreased.  The LISA and HEROS observations demonstrate that in 2018 the density of the expanding gas was still too high to display a significant increase in ionization and that the ejecta was barely beginning to "thin". Conversely, in 2019 the density of the ejecta must have decreased enough to become substantially ionized.

Although we cannot compare the relative intensities at the two 2019 epochs along the whole wavelength range of the FIES spectra because of the scattered light contamination (May spectrum) and color losses (July and possibly May spectra), we can compare them where their continuum matches, i.e. in the range 4600-5200 \AA. This is sufficient to establish that the He~II line ($\lambda$4686) intensity remained constant or slightly increased, that [O~III]$\lambda$5007 slightly decreased and more so H$\beta$, while [Fe~VI]$\lambda$5147,5177 and [Fe~VII]$\lambda$4942 remained constant.  
%The comparison between the May and Jul 2019 FIES spectra provides information about the ejecta and the ionizing source. 
The weakening of the hydrogen emission together with the constancy of the He~II and coronal lines, requires that the ionizing source is still active. 

\begin{figure}
\centering
\includegraphics[width=9cm]{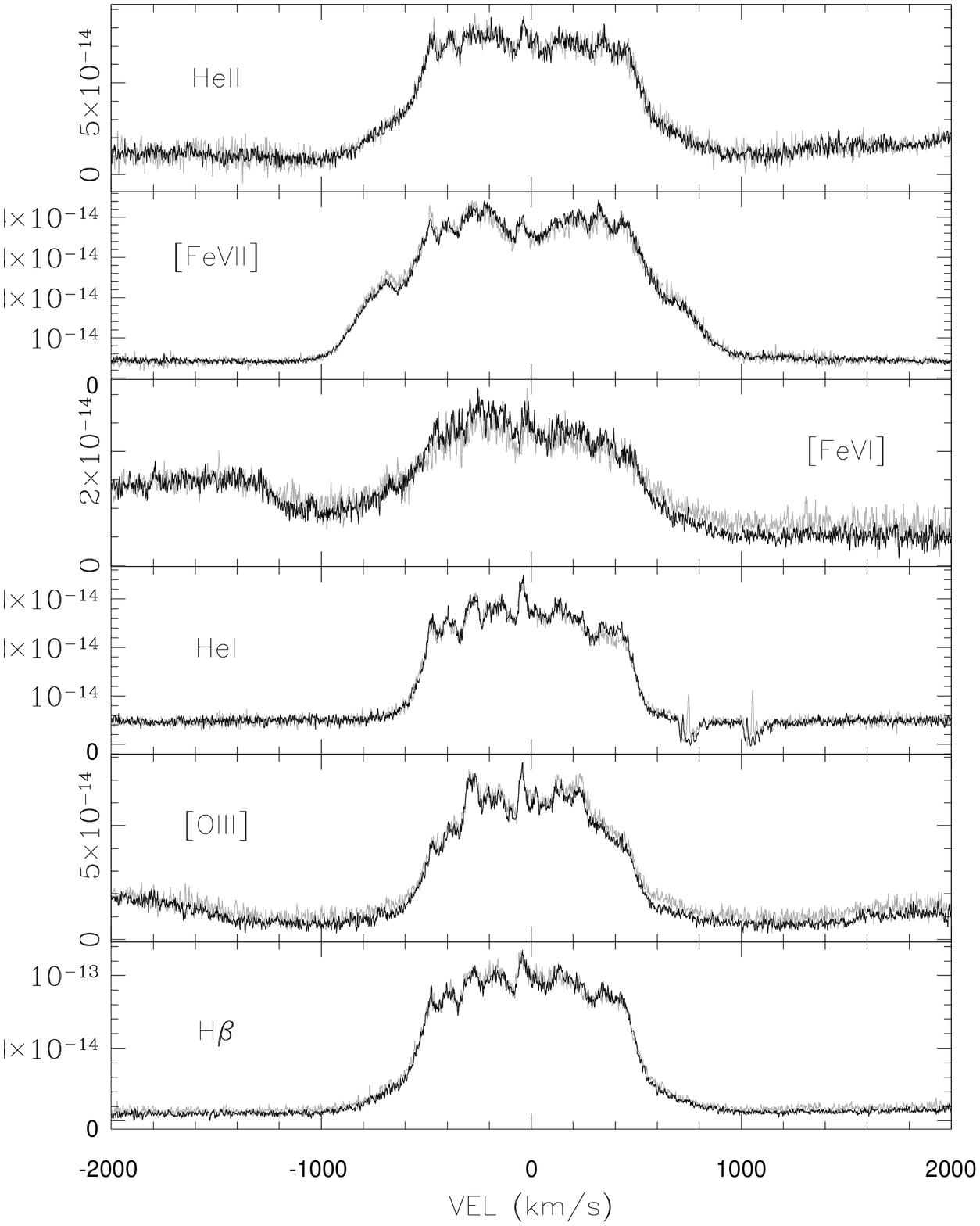}
\caption{Comparison of the coronal, permitted and nebular line profiles in the July 2019 FIES spectrum. The profiles share structures and width (see text for more details). We plot in grey the same profiles in the May 2019 FIES spectrum, showing invariance of the line profile evolution. Y-axes are flux in erg/cm$^2$/s/\AA.}
\label{profile}%
\end{figure}

The coronal lines emission profiles are at least as broad as the permitted and the nebular transitions. They also display the same structures and their profiles are unchanged between the two epochs (Fig.\ref{profile}). This precludes the role of shocks in their formation. Shocks excitation of different regions of the ejecta, or between the ejecta and any circumstellar material, should be localized at the colliding region and therefore display marked differences between coronal and permitted or nebular line profiles. They should also produce changing profile with time as the shock propagates.
That said, we can constrain the incident spectral distribution of the ionizing source using the coronal lines and transitions from elements that present a range of ionization, e.g. in addition to [Fe~VI], [Fe~VII] and possibly [Fe~X], He~I and He~II\footnote{The relative intensity of the [Ar~III] to [Ar~V] transitions cannot be compared because they lay in the region of scattered light.}. % as well [Ar~III] to [Ar~V]. 
The corresponding ionization potential (IP) are in the range 75 to 230 eV for iron, and 24 to 54 eV for helium. Since [Fe~X] is very weak and since we certainly do not detect [Fe~XIV] or [Fe~XI], we place the upper cut-off of the ionizing source in the range 200-250 eV (the IP for Fe$^{10+}$ is $\sim$260 eV). The stronger [Fe~VII] and [Fe~VI] transitions compared to the (putative) [Fe~X] (which has a transition probability $\sim$100 times greater) requires substantial flux from the ionizing source in the range 75-100 eV. It cannot be much below that otherwise He$^+$ would be over-ionized (although the fraction of doubly ionized helium is increasing). On the low energy side, we place the low cutoff at about 20 eV in order to have H just recombining (it weakens between May and July because of expansion) and He ionizing. %It is difficult to assess the relative intensity variation of the argon transitions because of the scattered light contamination in the May spectrum and their intrinsic weakness, but the [Ar~V] emission may have weakened less than [Ar~III] and [Ar~IV], suggesting energy peaking above 60 eV consistent with the inference from the Fe coronal lines. 

%To complete our analysis of the 2019 FIES spectra we note the following.  

\subsection{Ejecta mass and filling factor}\label{numbers}

\begin{figure}
\centering
\includegraphics[width=9cm]{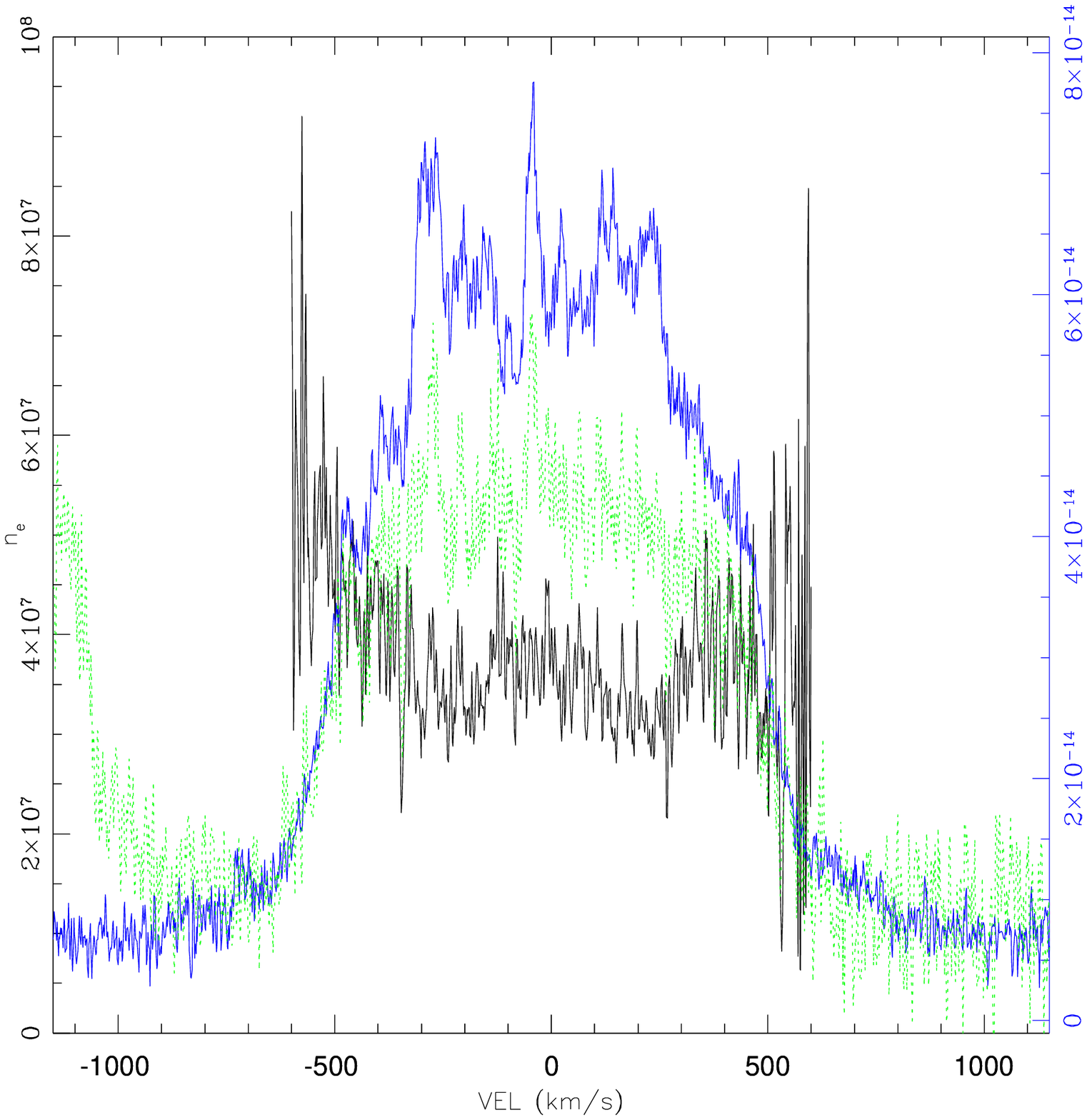}
\caption{The [O III] emission lines (July 2019 FIES spectrum) and the corresponding electron density in velocity space. The [O III]$\lambda$5007 is in blue solid color and the [O III]$\lambda$4363 is in green dotted color. Their units in erg/s/cm$^2$/\AA, is indicated in blue on the right axis. The black solid line is the  corresponding electron density, $n_e$, per resolution element (units in cm$^{-3}$ on the left axis). The $n_e$ is plot only within the velocity range [-600,+600] km/s, since beyond it, the noise dominates. 
%electron density (black) within the ejecta as determined from the July FIES spectrum. In gray color is the noisy portion of our density calculation.  We overplot in blue the profile of the [O~III] $\lambda$5007 emission. 
}
\label{oiii}%
\end{figure}

The invariant structures in the line profile (both forbidden, i.e. optically thin transitions by definition, and in the Balmer lines) indicate that the gas is in free expansion and optically thin. The observed decrease of the H$\beta$ emission (-15-20\%) in consistent, within the errors, with the density dropping with time, $t$, as $t^{-3}$ (-20-25\%) typical of freely expanding gas in ballistic motion and of CNe. Hence, we can constrain the electron density from the [O~III] lines as we have done for late nebular spectra of CNe (e.g. Shore et al. 2013a, 2013b; Shore et al. 2016; Mason et al. 2018). 
We are aware, however, that the relative intensity of the [O~III] components in 17hx indicates that collisional de-excitation is not negligible and the density must exceed the  critical density. 
We verified that for spectra of CNe whose [O~III] emission had similar relative strength to 17hx,  the derived ejecta mass is smaller than that computed from later spectra because of the collisional damping of the diagnostic lines. We therefore used the [O~III] diagnostic to estimate the nebular density and constrain the ejecta mass, being aware that the first is likely underestimated because of the wavelength dependent flux losses (see Section 2) and the latter will be underestimated because of collisional damping of the forbidden transitions. 
Fig.\ref{oiii} shows the nebular density derived using the  [O~III] diagnostic for an assumed electron temperature of T$_e\sim$10000~K (Osterbrock 1989, chapter 5), velocity bin per velocity bin. Since the 4959~\AA \ component is heavily blended with [Fe~VII] we disentangled its contribution using the 5007 \AA \ profile scaled by the ratio of their transition probability (2.9 in NIST). We find an average density of $\sim$3.5-4$\times$10$^7$ cm$^{-3}$ with some minor fluctuation across most of the line width. 

We constrained the ejecta's geometry by mimicking the emission line profiles with our bicone code, i.e. a Monte Carlo simulation that scatters,  in ballistic but otherwise random distribution, 10-30 thousand points (representing, each, individual ejecta structures) within a biconical geometry that can vary in opening angle, thickness and orientation (Shore et al. 2013; Mason et al. 2018). The closest matching model has maximum velocity v$_{max}$=1500 km/s, high inclination ($\sim$80 deg), wide opening angle ($\sim$140 deg) and is geometrically thin compared to other novae (i.e. v$_{min}\sim$50 to 60\% of v$_{max}$ compared to $\sim$30\% for nova Mon 2012, nova Del 2013 and nova Cen 2013, see Shore et al. 2013a, 2016, and Mason et al. 2018, respectively). 
With the density, its radial gradient and the geometry of the ejecta we can derive the ejecta mass upper limit for a filling factor of unity: M$_{ej}\simeq$9$\times$10$^{-3}$ M$_\odot$. %We emphasize that, although this estimate requires the departure from sphericity given by modeling, the effect is relatively small and, most importantly, it is independent of distance.   
We emphasize that, although this estimate is model dependent, requiring the departure from sphericity, it is independent of distance. 
We know, however, that the filling factor, $\varepsilon$, is $<$1 since the structures in the line profiles indicate fragmented ejecta.  The question is whether $\varepsilon$ can reduce the ejecta mass to the range of normal CNe. To obtain $\varepsilon$ we need the absolute luminosity of a purely recombination line, H$\beta$, sampling the same volume as the nebular diagnostic. This requires knowing the distance. The observed line of sight H~I 21 cm emission is optically thin and its matching profile with the atomic interstellar absorption lines (see Fig.\ref{lab21}) indicates that 17hx is fairly distant. 
Using the Galactic rotation curve, for the highest observed radial velocity of both the 21 cm emission and  Na~I and Ca~II absorptions, v$_{LSR}\simeq$120 km/s, we obtain a line of sight distance of $\approx$ 7.6 kpc.  Moreover, the total reddening inferred from infrared maps in 17hx direction is E(B-V)=1.5 mag  (Schlafly \&  Finkbeiner 2011) and is about twice our derived value (see Section 2.1). Considering that most of the dust is confined within the Solar circle, our derived E(B-V) is consistent with 17hx being located at about the same distance as the Galactic center.  We therefore take 7.6 and 8.5 kpc as a lower and upper limits for the distance to constrain $\varepsilon$. The dereddened H$\beta$ integrated flux  is L(H$\beta)_{obs}$=2.04$\times$10$^{-11}$  erg/cm$^2$/s/\AA \ (=1.95$\times$10$^{-12}$ in the original extinct spectrum).  As in our previous studies, we define $\varepsilon\equiv$L(H$\beta)_{obs}$/L(H$\beta)_{predicted}$ where the prediction is based on our computed $n_e$ and volume from the models, assuming case B recombination. The latter may overestimate the real recombination rate, given the high densities of the ejecta.  For the distance bounds, we obtain 0.08$\lesssim\varepsilon\lesssim$0.1, yielding ejecta mass in the range 7$\times10^{-4}\lesssim$M$_{ej}\lesssim$ 9$\times10^{-4}$ M$_\odot$.  These estimates are at least an order of magnitude higher than the largest values derived for CNe, especially those we have similarly analyzed.

%-----------------------------------------------------------------

\section{Discussion and conclusions}

ASASSN-17hx resembles a CN in some properties, but it is strikingly anomalous relative to the majority of CNe in a number of other aspects. 
Specifically, 17hx can be explained with the same dynamical and bolometric physical model we have advocated for CNe (Shore et al.  2012, 2013a, 2016, de Gennaro Aquino et al. 2014, Mason et al. 2018) that does not require a wind, or dynamically interacting multiple ejections. 
The light curve results from redistribution and reprocessing of the photons emitted from a central source as they pass through the ballistically expanding ejecta. The ejecta are structured as in CNe. The spectra developed, analogously to all CNe, from an initial opaque  phase, through a transition/semi-opaque phase, to a transparent nebular one. 
The oddities are in the details. 

The early outburst was dominated by large oscillations that for amplitude and interval between peaks are unusual for CNe. The nebular spectra displayed unusual line strength in the coronal emissions. The estimated ejecta mass is higher or much higher than typical CNe and this, combined with the observed velocity, implies a larger kinetic energy than typical CNe. % Finally, in light of the possible photometric amplitude of the outburst, that is not remarkable for a CN, the kinetic energy inferred for the ejecta is much greater. 
We now discuss each of these points in detail. 

\subsection{On the light curve oscillations}
Our early spectroscopy demonstrates that 17hx optically thick phase oscillations were accompanied by changes in the degree of ionization of the expanding gas, with higher ionization observed at the minima than at maxima. 
Thus, the photometric and spectroscopic changes can result from alternating ionization and recombination waves propagating through the expanding gas as the underlying source varied in brightness and SED (or hardness).  When the incident flux is higher or harder, the pseudo-photosphere recedes, the continuum peaks at shorter wavelength  the visible and the NIR fluxes drop, the iron peak elements ionize and He~I is excited. This is a photometric minimum. If the incoming radiation is softer, the pseudo-photosphere expands, the brightness of the continuum at long wavelength increases and the spectrum returns to its previous state of lower excitation and ionization degree in the emission line spectra. This is a maximum. Once the ejecta turns sufficiently transparent and the pseudo-photosphere disappears, there are no longer continuum oscillations, independent of the behavior of the underlying ionizing source. However, changes might be observed in the emission lines, as reported for, e.g., V1494 Aql (Iijima \& Esenoglu 2003) and V603 Aql (McLaughlin 1943). 
%Changes were recorded in the emission lines of other CNe (e.g. V1494 Aql in Iijima and Esenoglu 2003, V603 Aql in McLaughlin 1943) supporting a similar driver. 

The ability of a freely expanding ejecta to display photometric oscillations in response to an illuminating  variable source depends on the relative timescale of the source oscillations and radiative diffusion timescale of the expanding gas. The latter decreases as $\sqrt{\tau}$, where $\tau$ is the ejecta optical depth, which decreases as $time^{-2}$ with the expansion. 
If the expansion is very fast (i.e. expansion timescale $\ll$ variation timescale of the source) the opacity drops immediately and there will not be detectable  oscillations in the light curve. If instead, the source variations are much more rapid (i.e. high frequency pulses), than the diffusion timescale, the ejecta will act as low pass temporal filter and just display an average brightness or low amplitude fluctuations. When the source pulse intervals are comparable to the diffusion timescale, the ejecta will display oscillations. These might appear very slow when the density (i.e. $\tau$) is high, since the diffusion timescale is longer, but will become more frequent as the density (and therefore $\tau$) drops. The amplitude of the oscillations decreases with time since larger portions of the ejecta become transparent because of the expansion. 
This can explain both the 17hx light curve and the "O" (oscillation) and "J" (jittering) type light curves shown in Strope et al. (2010). 

This bring us to the question of the nature of the central source. The pulses from the underlying source must be sufficiently intense and rare to produce the observed effect on the ejecta. Smaller  amplitudes and more closely spaced peaks like those observed in, e.g., DQ Her or nova Sgr 2015b, would require higher pulse frequencies and lower amplitudes. The light curve jitters or oscillations could appear already at maximum or during early decline, depending on the initial ejecta opacity and the pulse cadence of the underneath source.  Conversely, the smoothly declining light curves would either have no pulses in the underlying source or a very rapidly expanding ejecta or a combination of the two. In any case, given the variety of observed jitters and oscillations the pulses should be similar to non-Gaussian noise with an occasional stronger signal capable of producing significant/outstanding variations in the light curve. These pulses are not the oscillations observed in the x-ray count rate at the "onset" of the supersoft source phase: those are likely too rapid, too small and, most importantly, explained by the differential changes in the transparency of individual intervening ejecta structures. The non-Gaussian noise-like pulses could result from the white dwarf (WD), or the resumption of accretion, or any temporary unstable interaction that might be occurring in the post-outburst phase on timescales of days to weeks. We suggest that those pulses, especially when they appear at maximum,  could be due to unstable burning by a nuclear source igniting just at the critical temperature. This depends on the mass of the WD. Massive WDs reach higher temperatures while mixing occurs on a relatively small envelope. Instead, low mass WDs reach lower temperatures closer to the critical value for the TNR ignitions so that, because of mixing, we can imagine a situation of marginal stability and  intermittent burning. The envelope mass is larger and mixing take longer and is less uniform. The burning layer would be embedded below a substantial envelope which would be partially radiative in its outer part, making the convection zone deeper and the supersoft source appear softer. The lower the WD mass, the more extreme this could be. The reduced degeneracy of the envelope would make it similar to unstable shell burning (Schwarzschild and Harm 1965; see also Jos\'e 2016, Iliadis 2007). 
Supporting this picture, the ionization degree displayed by the FIES spectra suggests a ionizing source that is  energetically limited to the range 20-250 eV and peaking between 75-100 or 75-150 eV. While we do not know its SED, it was not a SSS in its usual sense\footnote{For example, for Swift XRT spectra, the hardness ratio is usually estimated in the interval 0.2-1.0 keV, while our upper cutoff is at the lowest energy end of this band.}. Otherwise, more highly ionized Fe would have been detected. The inferred energy distribution is consistent with a buried nuclear source, or a cooling WD, or an accretion disk. 

\subsection{On the late spectral appearance and the coronal lines}
17hx is also peculiar for the relative strength of its coronal lines. When compared to classical novae followed until their "coronal phase" as in the CTIO survey by Williams et al. (1991 and 1994), we note the following. First, CNe\footnote{For the obvious reasons of the interaction between the ejecta with the donor's wind, symbiotic novae are excluded from this comparison.} that display coronal emissions are well into late nebular phase, with  [O~III] and [N~II] (typically much stronger than H$\beta$) being the dominant emission lines. Second, the lines of [Fe~VII] or higher transitions are always weak compared to both nebular and permitted transitions, usually  $\leq$0.2\,I(H$\beta$). Examples in the CTIO survey are nova Sco 1989b, nova Sct 1989, nova Cen 1991, nova Oph 1991a, nova LMC 1991, nova Sgr 1991 and nova Pup 1991 (Williams et al. 1991, 1994), and nova Aql 1999b (Iijima and Esenoglu 2003). The only objects that are similar to 17hx are nova Oph 1988 (V2214 Oph) in the CTIO survey, and V723 Cas, that was monitored by Iijima (2006) over 6 years. Both objects, like 17hx, developed strong coronal lines and weak [O~III]. They also displayed strong [Ne~V] lines, while for 17hx we lack such information as our spectra are not sufficiently extended in the blue. 
Both V2214 Oph and V723 Cas were slow novae with extreme oscillations during the early decline (whatever "early" might mean in this case). Strope et al. (2010) classified the first as an S-type nova (i.e. smooth light curve),  but they missed early decline data points which are published in Williams et al. (2003, although these lack error bar). V723 Cas, whose light curve Strope et al. (2010) assigned to the J class, had a remarkably slow spectroscopic evolution (Iijima  2006) that is very similar to 17hx. It took about 18 months to enter the optically thin phase. Two years after outburst it showed substantial spectroscopic evolution with dramatic strengthening of the He~II line $\lambda$4686 and the appearance of the coronal lines from [Fe~VI], [Fe~VII] and [Ca~V], while [O~III] remained much weaker than H$\beta$ (Iijima 2006). The ionization of the ejecta further increased, showing [Fe~X] over four years after outburst (Iijima 2006). In addition, Iijima (2006) reports broader line profiles in the higher ionization potential transitions (i.e. He~I and H) than those with low ionization potentials  (e.g. Fe~II) and similarly more extended wings of the coronal transition relative to the nebular lines in the late spectra. This matches the ionization structure observed in 17hx and is consistent with the ionization structure expected in a ballistically expanding gas powered by a central source.

\subsection{On the ejecta mass}

To continue the comparison of 17hx and V723 Cas we look at their ejecta masses. Iijima (2006) using integrated line fluxes derived an unusually large $n_e$, but a quite small mass (5$\times$10$^{-6}$ M$_\odot$), possibly incompatible with the observed slow evolution. The line fluxes reported in Iijima's Table 7 indicate a strongly collisionally damped [O III] ratio 4 years after outburst. 
We derived an independent value for the V723 Cas ejecta parameters, estimating the [O III] and H$\beta$ lines intensity from Iijima's Figure 14 and modeling their profiles with our Monte Carlo simulation and maximum expansion velocity  v$_{max}$=1600 km/s. The upper limit for the mass, for a filling factor of unity, is 0.03 M$\odot$. Adopting Schaefer's (2018) distance of 5.6$^{+1.9}_{-0.7}$ kpc we derive $\varepsilon$=0.01 and M$_{ej}$=3$\times$10$^{-4}$ M$_\odot$ for V723 Cas. Although uncertain these values are similar to those derived for 17hx. The two objects have a somewhat large ejecta mass, about an order of magnitude larger than typical CNe, and possibly more. The largest uncertainty is in the distance and, the closer the object, the more CNe-like the ejecta mass, but the smaller the filling factor which falls in the range for recurrent novae (e.g. T Pyx, Shore et al. 2013b).  It is interesting to note that these large ejecta masses expand with velocities that are comparable to those of typical CNe, implying larger kinetic energy.

In their nova sequences, Yaron et al. (2005) found an increase in the ejected mass with decreasing white dwarf mass, with a 0.4 M$_\odot$ WD ejecting $<7\times 10^{-4}$ M$_\odot$.  This is at the lower limit of our range for the ejecta mass in 17hx. Furthermore, since these were one dimensional simulations, the filling factor is unity by definition.  Hence, the mass and kinetic energies we derive are likely far higher than these models produced. The same holds for the TNR results from Starrfield et al. (2013), who found that even for low mass accretion rates and low mass WDs there is a TNR that can ultimately eject some mass, although neither as much as in Yaron et al. (2005) nor near our estimated value for 17hx. Starrfield et al. (2013) also remark that unsteady nuclear burning would result for accretion of solar composition material and no mixing. %We might notice that although just on the basis of the relative strength of the emission lines, 17hx abundances might be consistent with the solar ones.   

\subsection{On the progenitor}
Saito (2017) identified the 17hx progenitor with a Ks$\sim$16.7 mag star in the VVVX survey, also matching a nearby Gaia DR2 source of G=19.3 mag. The latter, once dereddened and scaled to 7.6 kpc (see also Evans et al. 2018), has an absolute magnitude M$_V\sim$3 mag, consistent with old nova absolute magnitudes determined by Selvelli and Glimozzi (2019). The same exercise repeated in K band produce M$_K$=2 mag which is somewhat bright for a cool main sequence companion, but compatible with an evolved donor.  Saito's identification implies that 17hx outburst amplitude was 11 mag. We note that there are not other visible objects nearby 17hx and that its outburst amplitude would be much larger than 11 mag should the identification not be confirmed. In, contrast, Gaia DR2 magnitudes and Schaefer's distance for V723 Cas indicate an absolute magnitude M$_V\lesssim$0.1 mag which does not match an ordinary old nova or quiescent cataclysmic variable.

%Should Saito's identification not be confirmed, the 17hx outburst amplitude was $>$11 mag. 
%In contrast, Gaia DR2 parallax\footnote{Note that the Gaia DR2 parallax for V723 Cas is only a 3$\sigma$ detection, possibly suggesting a lower limit to the distance.} and magnitudes of V723 Cas remnants indicate an absolute magnitude M$_V$=-0.58 mag, which does not match an ordinary old nova or quiescent cataclysmic variable. 

In conclusion, after this long presentation and physical dissection of the source properties using every tool we have developed for the study of CNe, we cannot be confident about the transient progenitor.  Although ASASSN-17hx has many of the earmarks of a classical nova, we may be dealing with an impostor.

\begin{acknowledgements}
EM deeply thanks John Telting at the NOT telescope for his prompt help, suggestions and exhaustive explanations. The authors are grateful to the people of the ARAS association for making freely available the spectra of their observations. In particular the authors thanks observers Umberto Sollecchia, Lorenzo Franco, Olivier Garde, Tim Lester and Joan Guarro Flo whose data has been displayed in the present paper. We thanks the observers registered to the AAVSO database who contributed to the photometric monitoring of ASASSN-17hx. We also thanks Kim Page for the help with the XRT data reduction and acknowledge Swift PI and operation staff for approving and planning the Swift observations. PK acknowledges support by the UK Space Agency. We also thank the referee, Nye Evans, for his precious feedback. 
\end{acknowledgements}

%%\bibliographystyle{aa} % style aa.bst
%%\bibliography{17hx_pap_reference.bib} % your references Yourfile.bib

\end{document}